\documentclass{jfm}

\usepackage{psfig}
\usepackage[dvips]{graphicx}
\usepackage{psfrag}
\usepackage{rotating}
\usepackage{natbib}
\title{Inertial waves in a rotating  spherical shell: attractors and
asymptotic spectrum}
\author[M. Rieutord, B. Georgeot and L. Valdettaro]{
M. RIEUTORD$^{1,2}$, B. GEORGEOT$^3$ \and L. VALDETTARO$^{1,4}$
}

\affiliation{
$^1$Observatoire Midi-Pyr\'en\'ees,
14 av. E. Belin, F-31400 Toulouse, France\\
$^2$Institut Universitaire de France\\
[\affilskip]
$^3$Laboratoire de Physique Quantique IRSAMC, Universit\'e Paul Sabatier,
118, Route de Narbonne F-31062 Toulouse Cedex 4, France\\
$^4$Dipartimento di Matematica, Politecnico di Milano, Piazza L.
da Vinci, 32, 20133 Milano, Italy
}

\pubyear{2000}
\volume{000}
\pagerange{000--000}
\date{\today}
\def\bbbr{{\rm I\!R}} 
\def\bbbn{{\rm I\!N}} 
\def\bbbq{{\mathchoice {\setbox0=\hbox{$\displaystyle\rm
Q$}\hbox{\raise
0.15\ht0\hbox to0pt{\kern0.4\wd0\vrule height0.8\ht0\hss}\box0}}
{\setbox0=\hbox{$\textstyle\rm Q$}\hbox{\raise
0.15\ht0\hbox to0pt{\kern0.4\wd0\vrule height0.8\ht0\hss}\box0}}
{\setbox0=\hbox{$\scriptstyle\rm Q$}\hbox{\raise
0.15\ht0\hbox to0pt{\kern0.4\wd0\vrule height0.7\ht0\hss}\box0}}
{\setbox0=\hbox{$\scriptscriptstyle\rm Q$}\hbox{\raise
0.15\ht0\hbox to0pt{\kern0.4\wd0\vrule height0.7\ht0\hss}\box0}}}}
\def\bbbt{{\mathchoice {\setbox0=\hbox{$\displaystyle\rm
T$}\hbox{\hbox to0pt{\kern0.3\wd0\vrule height0.9\ht0\hss}\box0}}
{\setbox0=\hbox{$\textstyle\rm T$}\hbox{\hbox
to0pt{\kern0.3\wd0\vrule height0.9\ht0\hss}\box0}}
{\setbox0=\hbox{$\scriptstyle\rm T$}\hbox{\hbox
to0pt{\kern0.3\wd0\vrule height0.9\ht0\hss}\box0}}
{\setbox0=\hbox{$\scriptscriptstyle\rm T$}\hbox{\hbox
to0pt{\kern0.3\wd0\vrule height0.9\ht0\hss}\box0}}}}
\newcommand{\tv}{ \rightarrow}

\newcommand{\beq}{\begin{equation}}
\newcommand{\beqa}{\begin{eqnarray*}}
\newcommand{\beqan}{\begin{eqnarray}}
\newcommand{\greq}{\begin{equation}\left. \begin{array}{l}}
\newcommand{\egreq}{\end{array}\right\} \end{equation}}
\newcommand{\nngreq}{\[\left\{ \begin{array}{l}}
\newcommand{\nnegreq}{\end{array}\right. \]}
\newcommand{\egreqn}[1]{\end{array}\right\} \label{#1}\end{equation}}
\newcommand{\eeq}{\end{equation}} 
\newcommand{\eeqn}[1]{\label{#1}\end{equation}} 
\newcommand{\eeqa}{\end{eqnarray*}}
\newcommand{\eeqan}[1]{\label{#1}\end{eqnarray}}

\newcommand{\noi}{ \noindent }
\newcommand{\ie}{{\it i.e.}\ }

\newcommand{\lp}{ \left(}
\newcommand{\rp}{ \right)}
\newcommand{\lc}{ \left[}
\newcommand{\rc}{ \right]}

\newcommand{\na}{ \vec{\nabla} }
\newcommand{\lap}{ \nabla^2 }

\newcommand{\cth}{ \cos\theta }
\newcommand{\sth}{ \sin\theta }

\newcommand{\vu}{\vec{u}}

\newcommand{\er}{\vec{e}_r}
\newcommand{\es}{\vec{e}_s}

\newcommand{\ez}{\vec{e}_z}
\newcommand{\epar}{\vec{e}^{\,\parallel}}
\newcommand{\eper}{\vec{e}^\perp}

\newcommand{\vk}{\vec{k}}

\newcommand{\vn}{\vec{n}}

\newcommand{\vv}{\vec{v}}
\newcommand{\vV}{\vec{V}}

\newcommand{\demi}{\frac{1}{2}}

\newcommand{\od}[1]{\mbox{${\cal O}(#1)$}}
\newcommand{\dq}[1]{\frac{\partial  #1}{\partial q}}

\newcommand{\ds}[1]{\frac{\partial  #1}{\partial s}}
\newcommand{\dt}[1]{\frac{\partial  #1}{\partial t}}

\newcommand{\deta}[1]{\frac{\partial #1}{\partial \eta}}
\newcommand{\dmu}[1]{\frac{\partial #1}{\partial \mu}}
\newcommand{\dup}[1]{\frac{\partial  #1}{\partial u_+}}
\newcommand{\dum}[1]{\frac{\partial  #1}{\partial u_-}}
\newcommand{\dcp}{d{\cal C}(\Psi)}

\newcommand{\dxi}[1]{\frac{\partial  #1}{\partial \xi}}

\newcommand{\dx}[1]{\frac{\partial  #1}{\partial x}}
\newcommand{\dy}[1]{\frac{\partial  #1}{\partial y}}
\newcommand{\dY}[1]{\frac{\partial  #1}{\partial Y}}
\newcommand{\dz}[1]{\frac{\partial  #1}{\partial z}}

\newcommand{\dupdum}[1]{\frac{\partial^2  #1}{\partial u_+\partial u_-}}
\newcommand{\ddx}[1]{\frac{\partial^2  #1}{\partial x^2}}
\newcommand{\ddy}[1]{\frac{\partial^2  #1}{\partial y^2}}

\newcommand{\dddY}[1]{\frac{\partial^3  #1}{\partial Y^3}}

\newcommand{\dsdz}[1]{\frac{\partial^2  #1}{\partial s\partial z}}
\newcommand{\dds}[1]{\frac{\partial^2  #1}{\partial s^2}}
\newcommand{\ddz}[1]{\frac{\partial^2  #1}{\partial z^2}}

\newcommand{\dsint}[1]{\frac{1}{\sin\theta}\frac{\partial}{\partial\theta}(\sin\theta #1)}

\newcommand{\dphi}[1]{\frac{\partial  #1}{\partial \phi}}
\newcommand{\dnphi}[1]{\frac{d #1}{d \phi}}

\newcommand{\disp}[1]{\displaystyle #1}
\newcommand{\eq}[1]{(\ref{#1})}

\newcommand{\supapp}{\raisebox{-.7ex}{$\stackrel{>}{\sim}$}}
\def\og{\leavemode\raise.3ex\hbox{$\scriptscriptstyle\langle\!\langle$}}
\def\fg{\leavemode\raise.3ex\hbox{$\scriptscriptstyle\rangle\!\rangle$}}

\begin{document}
\bibliographystyle{jfm}
\maketitle
%
%

\begin{abstract}
We investigate the asymptotic properties of inertial modes confined in
a spherical shell when viscosity tends to zero. We first consider the
mapping made by the characteristics of the hyperbolic equation
(Poincar\'e's equation) satisfied by inviscid solutions.
Characteristics are straight lines in a meridional section of the
shell, and the mapping shows that, generically, these lines converge
towards a periodic orbit which acts like an attractor (the associated
Lyapunov exponent is always negative or zero). We show that these
attractors exist in bands of frequencies the size of which decreases
with the number of reflection points of the attractor. At the bounding
frequencies the associated Lyapunov exponent is generically either zero
or minus infinity. We further show that for a given frequency the
number of coexisting attractors is finite.

 We then examine the relation between this characteristic path and
eigensolutions of the inviscid problem and show that in a purely
two-dimensional problem, convergence towards an attractor means that
the associated velocity field is not square-integrable. We give
arguments which generalize this result to three dimensions. Then, using
a sphere immersed in a fluid filling the whole space, we study the
critical latitude singularity and show that the velocity field diverges
as $1/\sqrt{d}$, $d$ being the distance to the characteristic grazing
the inner sphere.

We then consider the viscous problem and show how viscosity transforms
singularities into internal shear layers which in general betray
an attractor expected at the eigenfrequency of the mode.
Investigating the structure of these shear layers, we find that they are
nested layers, the thinnest and most internal layer scaling
with $E^{1/3}$-scale, $E$ being the Ekman
number; for this latter layer, we give its analytical form and show its
similarity to vertical $\frac{1}{3}$-shear layers of steady flows.
Using an inertial wave packet traveling around an attractor, we give a
lower bound on the thickness of shear layers and show how
eigenfrequencies can be computed in principle. Finally, we show that as
viscosity decreases, eigenfrequencies tend towards a set of values
which is not dense in $[0,2\Omega]$, contrary to the case of the full
sphere ($\Omega$ is the angular velocity of the system).

Hence, our geometrical approach opens the possibility of describing the
eigenmodes and eigenvalues for astrophysical/geophysical Ekman numbers
($10^{-10}-10^{-20}$), which are out of reach numerically, and this for a
wide class of containers.
\end{abstract}

\section{Introduction}

Inertial waves, which propagate in rotating fluids thanks to the
restoring action of the Coriolis force, can generate very singular fluid
flows when they are confined in a closed container. These very special
properties of  inertial modes were first noticed in the theoretical
work of K. Stewartson and others
\cite[]{StRic69,stewar71,stewar72a,stewar72b,walt75b,LS79}. They
appeared again recently in numerical investigations by \cite{HK95},
\cite{rieu95}, \cite{RV97}, \cite{FH98} and show an even greater
generality since they are also present in stratified fluids
\cite[]{ML95,RN99} or rotating stratified fluids \cite[]{DRV99}.

The particularity of all these waves (inertial, gravity,
gravito-inertial) is that their associated modes are solutions
of an ill-posed boundary-value problem when they are confined in 
a close container: the partial differential equation is of hyperbolic or
mixed type
in the spatial variables. This yields all kinds of singularities. When
viscosity is included, these singularities are regularized but they still
play a central role in featuring the shape of inertial modes of a rotating
spherical shell; in particular, they control the asymptotic limit of
small diffusivities which is the relevant limit for astrophysical or
geophysical applications.

The aim of this paper is to present what we believe to be the
asymptotic limit of inertial modes in a spherical shell when viscosity
tends to zero. In the first part of the paper we shall present the main
features of the solutions of this problem when viscosity is omitted. For
this purpose we examine the trajectories of characteristics in a
meridional plane of the shell as if they were trajectories of a
dynamical system in some configuration space. We then focus on the relation
between these trajectories and the eigenfunctions in two and three
dimensions. We end this part with a close look at the critical latitude
singularity. In the second part we investigate the changes brought on by
viscosity and we examine more closely the structure of shear layers
which arise. Then, by studying the behaviour of a wave-packet, we
show how eigenvalues and eigenmodes may be computed in the asymptotic
limit of a small viscosity. We conclude this part by a brief
discussion of the distribution of eigenvalues in the complex plane.
The paper ends with a discussion of the more general cases
including containers with a different shape
and of the applications of the present theoretical results.

	As this paper is rather long and goes through some mathematical
developments which may be skipped at first reading, we suggest the
casual reader to skip subsections \S 2.2.1-5, 2.3.1-2 and 2.4.1-2
and be lead by the introductions of sections 2.2, 2.3 and 2.4 and then
jump to 2.5 and 2.6. The second part is not so mathematical but the
details of the boundary layer analysis (\S 3.2.2-4) can be skipped at
first reading.

\section{Some properties of inviscid solutions}

\subsection{Equations of motion}

We consider a fluid with no viscosity contained in a spherical shell
whose outer radius is $R$ and inner radius $\eta R$ with $\eta<1$. The
fluid is rotating around the $z$-axis with the angular velocity
$\Omega$. Using $(2\Omega)^{-1}$ as the time scale and $R$ as the length
scale, small amplitude perturbations obey the linear equation

\greq
\dt{\vu} + \ez \times \vu = -\na p \\
\\
\na\cdot\:\vu=0
\egreqn{eqmo}

\noi where $\vu$ is the velocity field of the perturbations
and $p$ is the reduced pressure perturbation. The boundary conditions
are simply

\beq \vu\cdot\er = 0 \qquad {\rm at}\quad r=\eta \quad{\rm and}\quad r=1
\eeq

\noi As in \cite{RV97} and \cite*{DRV99} we shall use spherical
coordinates $(r,\theta,\varphi)$ or cylindrical coordinates
$(s,\varphi,z)$. $\vec{e}_q$ will denote the unit vector associated with
the coordinate $q\in \{r,\theta,\varphi,s,z\}$.

When the time dependence of the solutions is assumed proportional to
$\exp(i\omega t)$, equations \eq{eqmo} may be cast into a single
equation for the pressure, namely

\beq \lap p - \frac{1}{\omega^2}\ddz{p} = 0 \eeq

\noi which has been referred to as Poincar\'e equation since the work of
\cite{Cartan22}.
This equation is completed by the boundary condition $\vu\cdot\er=0$
which reads

\beq -\omega^2\er\cdot\na p +i\omega(\ez\times\er)\cdot\na p +
(\ez\cdot\er)(\ez\cdot\na p) =0 \eeq

\noi when expressed with the pressure; it applies at $r=\eta$ and $r=1$.

As is well-known \cite[]{Green69}, the Poincar\'e equation is hyperbolic
since for all modes $\omega < 1$. Therefore, a first step in the
analysis of this equation is the determination of the characteristics
surfaces; for this purpose, we note that second-order derivatives of this
equation read

\[ \ddx{p} + \ddy{p} - \frac{\alpha^2}{\omega^2}\ddz{p} \]

\noi where $\alpha^2=1-\omega^2$. $x$ and $y$ are the cartesian coordinates
in a plane $z=0$. These second-order derivatives define characteristics surfaces
such as $z-f(x,y)=0$ where $f$ verifies~:

\beq \lp\dx{f}\rp^{\!2} + \lp\dy{f}\rp^{\!2} =
\frac{\alpha^2}{\omega^2}\, .\eeqn{sfc}

\noi These surfaces are known as `surfaces of constant slope' as any
tangent plane makes the same angle $\vartheta=\frac{\pi}{2} - \arcsin\omega$
with the equatorial plane $z=0$. In fact, these surfaces may be
generated by a family of such planes or by cones which aperture angle is

\beq\lambda=\arcsin\omega\eeq

\noi which is also the critical latitude in a sphere.  In a meridional
plane, the trace of these surfaces are simply straight lines making the
angle $\lambda$ with the rotation axis.

A second step in the analysis of the Poincar\'e equation is to examine
the separability of the variables. Because of the symmetry of the problem
with respect to rotations around the $z$-axis, the $\varphi$ variable may
always be separated from the two others. This implies that solutions may
always be expressed as

\[ \sum_{m} p_m(r,\theta) e^{im\varphi} \]

\noi and that each Fourier component $p_m(r,\theta)$ is independent of
the others.

The two other coordinates, however, are not separable in the general
case. This point may be understood easily if we recall that through a
linear transformation of the $z$-coordinate, the Poincar\'e equation may
be transformed into the Laplace equation as first shown by
\cite{Bryan1888} \cite[see also][]{Green69}. In this transformation the
boundaries transform into one-sheet hyperboloids. One therefore adopts
an ellipsoidal coordinate system within which one of the boundaries
is a surface of coordinate; unfortunately, since the two bounding
hyperboloids are
not confocal, coordinates can be only separated on one of the boundaries
of the spherical shell.
Hence, one may simplify, and actually solve, the problem either in the
full sphere \cite[see][]{Green69} or in the infinite fluid outside a
sphere (see below).

We have now seen the basic ingredients which make this problem
difficult: hyperbolicity (ill-posedness) and non-separability.

\subsection{Orbits of characteristics}\label{dyn_char}

As shown by the foregoing discussion the difficulty of the problem lies
in the behaviour of the solutions with respect to the coordinates in a
meridional plane $(s,z)$ or $(r,\theta)$. We shall therefore restrict
our analysis to this plane where characteristic surfaces are simply straight
lines; thanks to the $\exp(im\varphi)$-dependence, our results will apply
equally to axisymmetric or non-axisymmetric modes. Indeed, using
the separation of the $\varphi$-variable, we eliminate second-order
derivatives in $\varphi$ and characteristic surfaces are just cones
independent of $m$.
We shall therefore study, in the following subsections, the trajectories
of characteristics in a meridional plane as we did in \cite{DRV99}. Since
this is a rather technical matter, the reader may first skip it and
directly jump to section 2.3 where the results are summarized.

\subsubsection{The mapping}

From
\eq{sfc} we derive the well-known equations of the two families of
characteristics:

\beq \omega z \pm \alpha s = u_\mp \eeqn{charac}

\noi where $u_\pm$ will designate the characteristics coordinates. $u_+$
and $u_-$ are  constant along characteristics of positive and 
negative slopes, respectively.

\begin{figure}
\begin{minipage}[c]{0.5\textwidth}
\centerline{\includegraphics[width=7cm,angle=0]{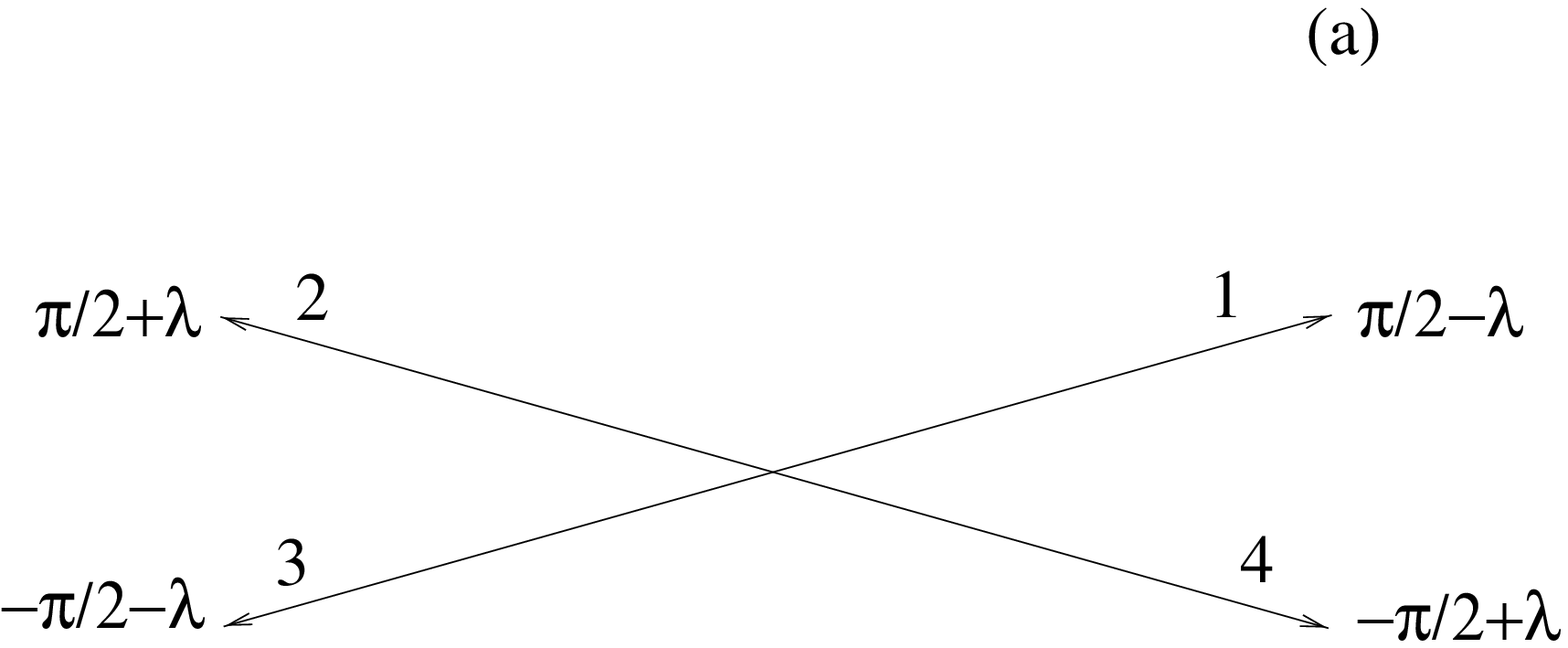}}
\end{minipage}\begin{minipage}[c]{0.5\textwidth}
\includegraphics[width=7cm,angle=0]{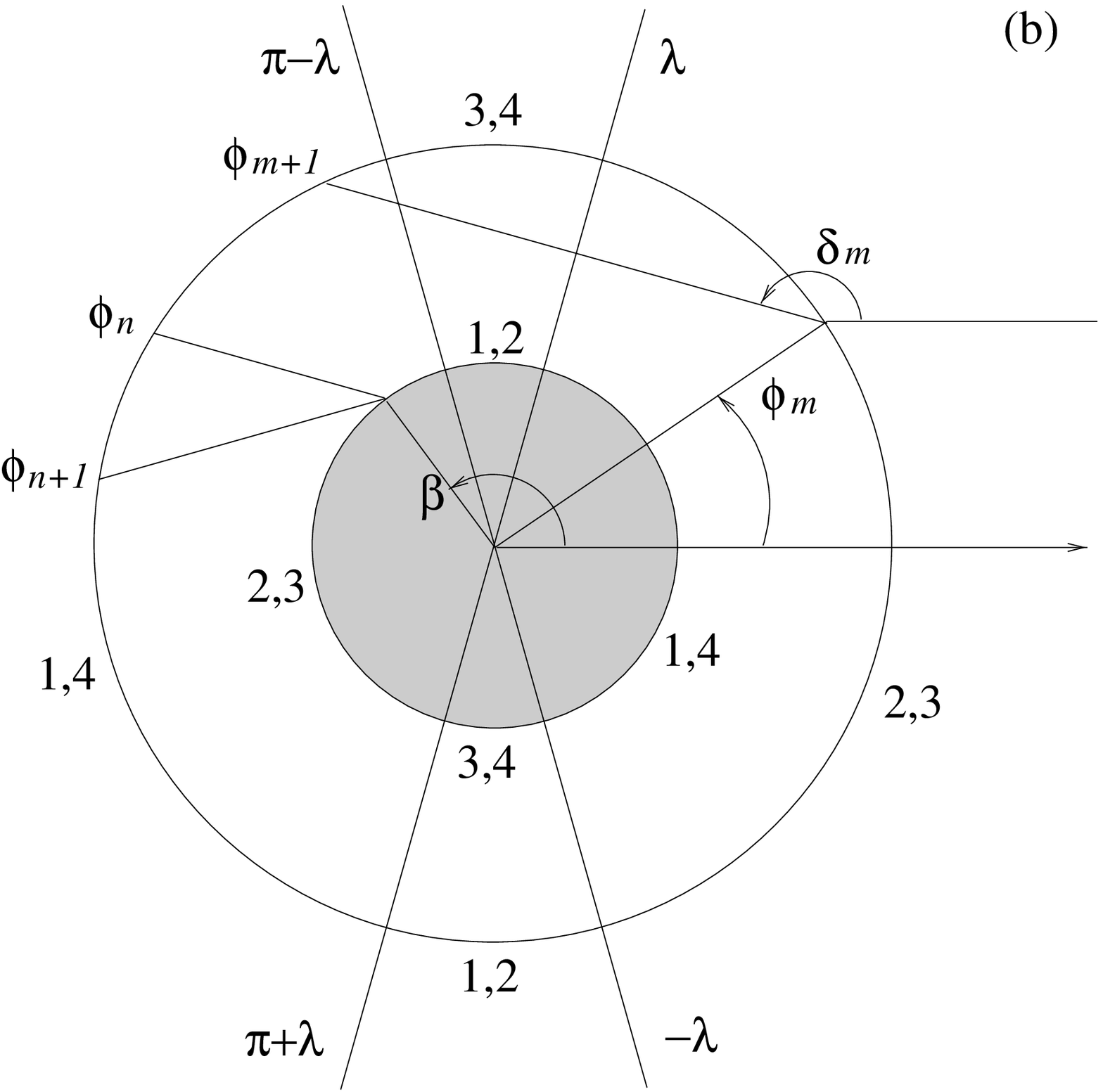}
\end{minipage}
\caption[]{(a) The four directions of propagation and the
corresponding four values of $\delta_n$'s.
(b) A sketch of the notations used to describe the mapping. The numbers
(1,4), (2,3) etc. indicate the possible directions of propagation of
characteristics as shown in the panel (a).}
\label{fig_rec}
\end{figure}

To describe the paths followed by characteristics, we first study the
map which relates the position of the n+1$^{\rm th}$
reflection point to the one of the n$^{\rm th}$ both taken on the outer
sphere. For this purpose we mark out these points by an angle
$\phi\in[0,2\pi]$ which is identical to the latitude when
$0\leq\phi\leq\pi/2$; this is indeed more convenient than the
colatitude. If one reflection is needed on the inner sphere then we have 

\greq
\sin(\phi_n-\delta_n) = \eta\sin(\beta-\delta_n) \\
\sin(\phi_{n+1}-\delta_{n+1}) = \eta\sin(\beta-\delta_{n+1})\\
\egreqn{cart1}

\noi where $\beta$ is the position of the reflection point on the inner
shell and $\delta_n=\pm\pi/2\pm\lambda$ is the direction of the
characteristic which can take four values as illustrated in
figure~\ref{fig_rec}a. If the two reflection points are simply connected
by one segment of characteristics then the recurrence relation is
either

\beq \phi_m + \phi_{m+1} = -2\lambda \; [2\pi] \eeqn{cart2}

\noi when the characteristic has a positive slope, or

\beq \phi_m + \phi_{m+1} = 2\lambda \; [2\pi] \eeqn{cart3}

\noi when the characteristic has a negative slope. These notations are
summarized in figure~\ref{fig_rec}b.

From the expression \eq{cart1}, \eq{cart2} and \eq{cart3} one can
compute the map 

\[ \phi_{n+1}=\left\{ \begin{array}{c} f_+(\phi_n)\\ f_-(\phi_n)
\end{array}\right. \]

\begin{figure}
\centerline{
\includegraphics[width=7cm,angle=0]{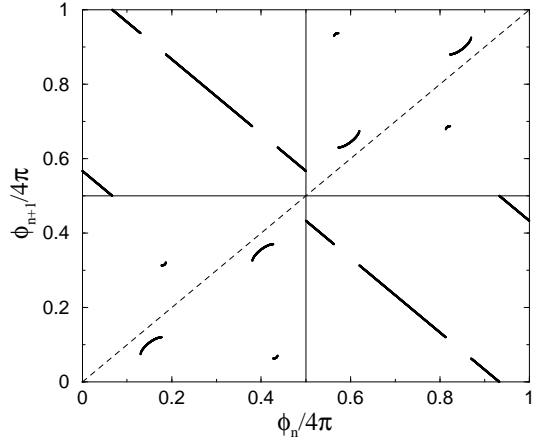}}
\caption[]{ The
resulting mapping in the case of a shell with $\eta=0.35$ when the
frequency is $\omega=0.40782$. Twelve points of discontinuity indicate
the projection of the `shadow' and the critical latitude of the inner
sphere on the outer one (see text); the four apparent discontinuities due
to the periodicity in the [0,4$\pi$] interval are not counted.}
\label{the_mapping}
\end{figure}

 \noi Such a map is bi-valued since one may compute the
image by first applying a positive- ($f_+$) or a negative- ($f_-$)
slope characteristic. However, such a representation is not convenient
for iterating the map since for each iteration one has to decide
whether to use $f_+$ or $f_-$. We therefore constructed a single-valued
map which is defined in the following way: Considering a point
of the outer sphere, we mark it out by the angle $\phi\in[0,2\pi]$ if it
is to be iterated by $f_-$ or by the same angle plus $2\pi$ if it is to be
iterated by $f_+$.
We thus define the map:

\beqan f\,:\,  &[0,4\pi]&\longrightarrow [0,4\pi] \nonumber \\
          &\phi_n& \longrightarrow  f(\phi_n)=\phi_{n+1}
\eeqan{applic}

\noi which is one-to-one except at some points of discontinuity and
which can be easily iterated. An example of this map is
given in figure~\ref{the_mapping}.

One of the remarkable features of this
map is that it is not continuous. The discontinuities occur at the
colatitudes (in the first quadrant)

\[ \theta_\pm = \lambda \pm\arcsin\eta, \qquad
\theta_s=\lambda+\arcsin(\eta\cos2\lambda) \]

\begin{figure}
\centerline{\includegraphics[width=7cm,angle=0]{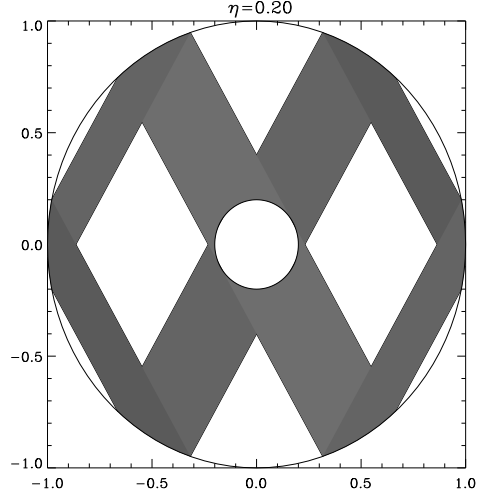}}
\caption[]{The shadow pattern for a spherical shell with $\eta=0.2$ when
$\lambda=\pi/6$ ($\omega=0.5$).}
\label{fig_shad}
\end{figure}

\noi $\theta_\pm$ are delineating the `shadow' projection of the inner shell
on the outer shell (see figure~\ref{fig_shad} for an illustration of the
shadow). They illustrate the case when a characteristic is
tangent to the inner sphere at critical latitude. $\theta_s$ is the
colatitude of the projection of the inner sphere's critical latitude on
the outer sphere. These angles ($\theta_\pm$) delimit
the intervals where the map is contracting $|f'|<1$ or dilating
$|f'|>1$ or neutral $|f'|=1$.

\begin{figure}
\centerline{\includegraphics[width=6.5cm,angle=0]{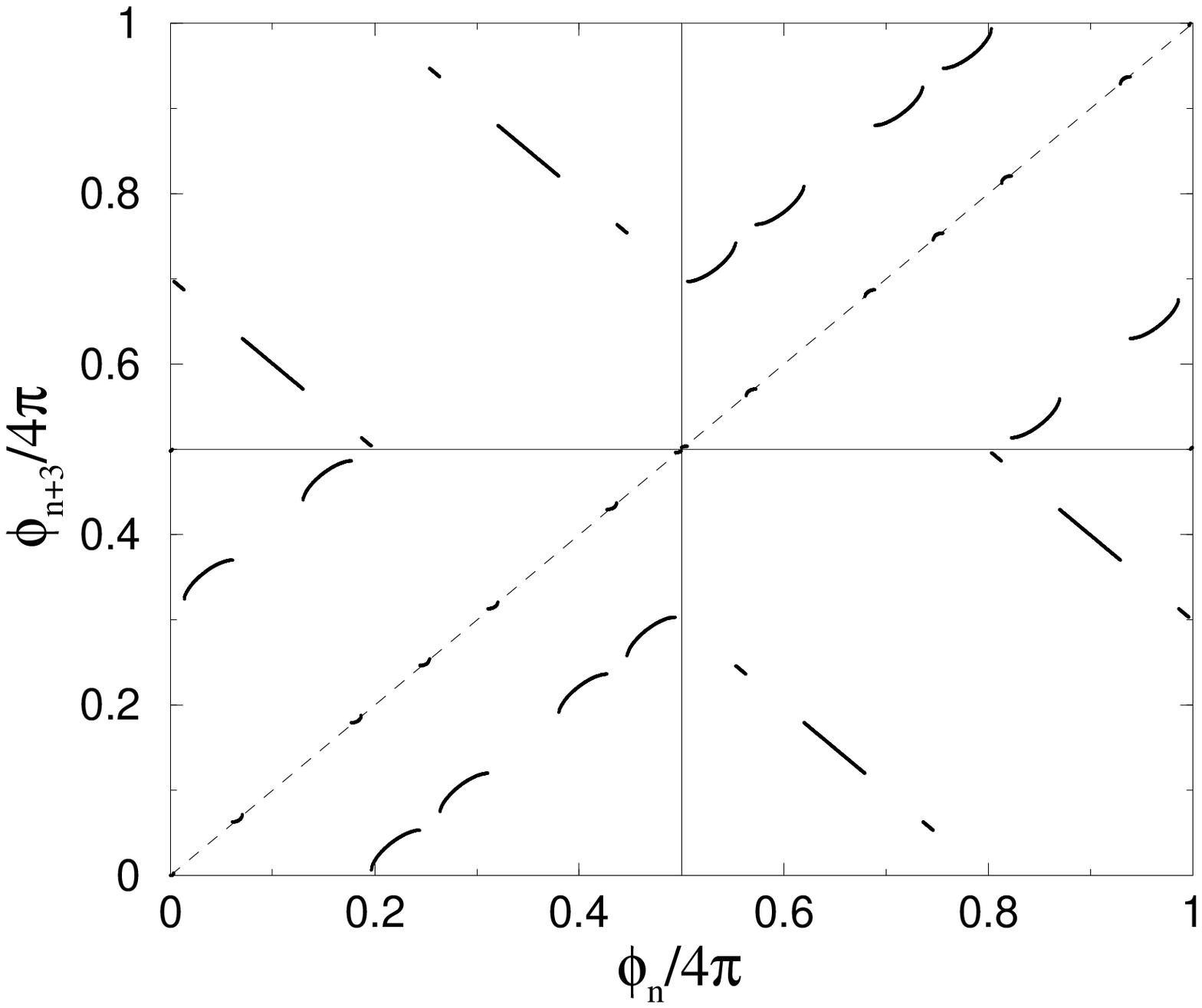}
\includegraphics[width=6.5cm,angle=0]{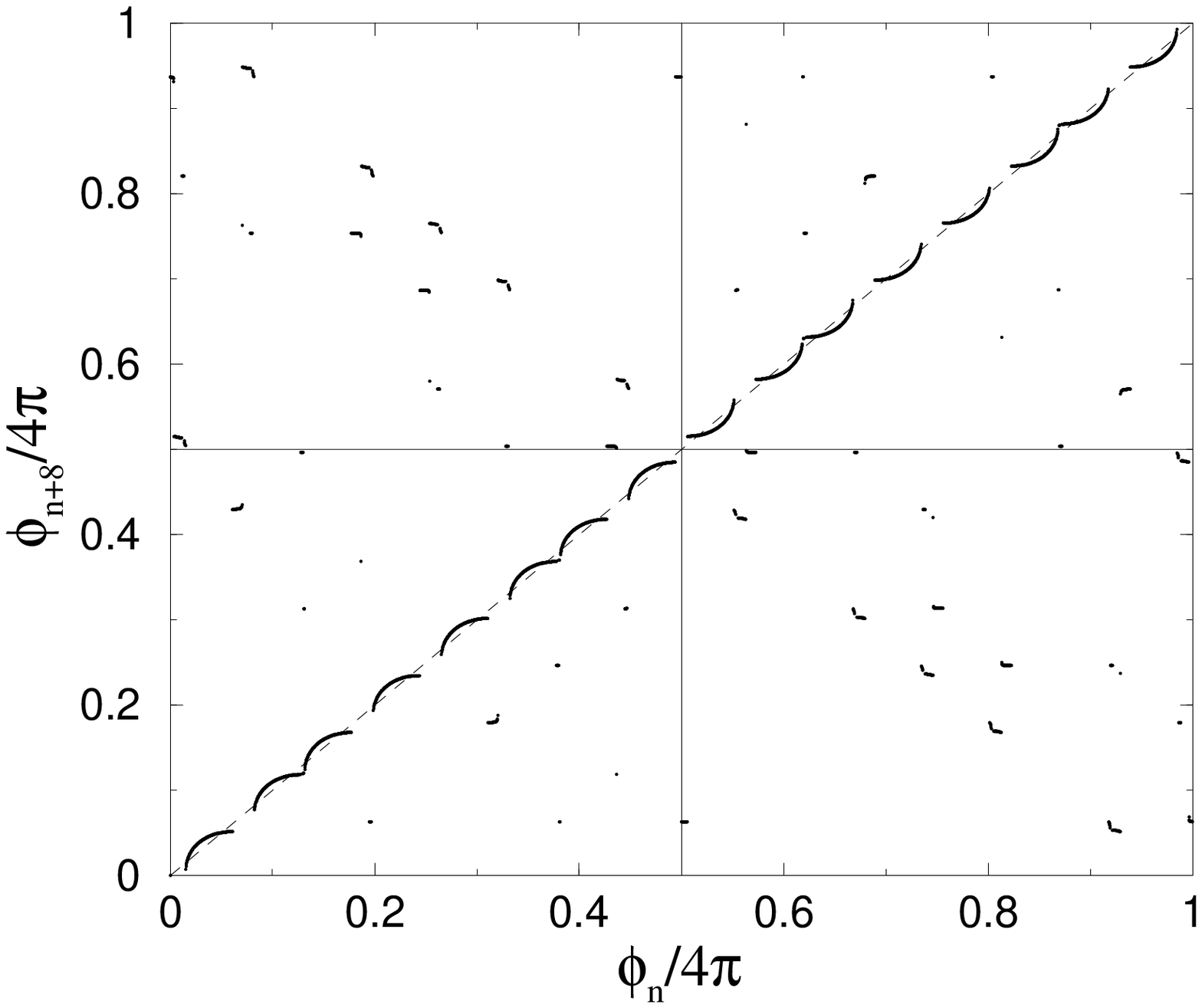}}
\caption[]{The third (left panel) and eighth (right panel) iterate of
the map in the case of a mode with $\omega=0.40782$ for a shell
of aspect ratio $\eta=0.35$: note the appearance of fixed points marked
out by the intersection of the curve with the line $\phi_{n+p}=\phi_n$.
They indicate the existence of attractive periodic orbits of period 3 and
8 respectively. Note also the increased number of discontinuities of
the mapping.}\label{fig_map} \end{figure}

\begin{figure}
\begin{minipage}[c]{0.5\textwidth}
\centerline{\includegraphics[width=.95\textwidth]{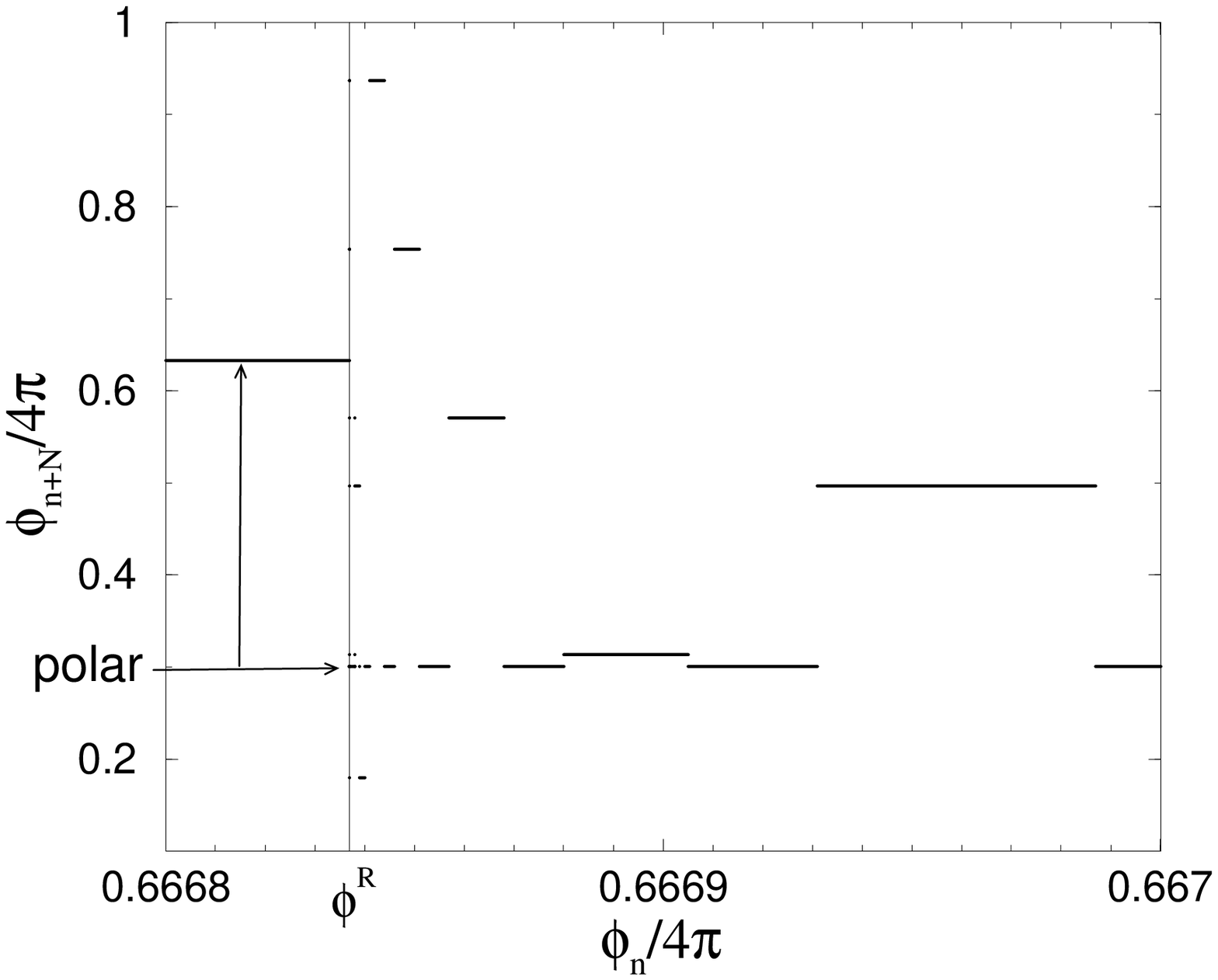}}
\end{minipage}\begin{minipage}[c]{0.5\textwidth}
\centerline{\includegraphics[width=.95\textwidth]{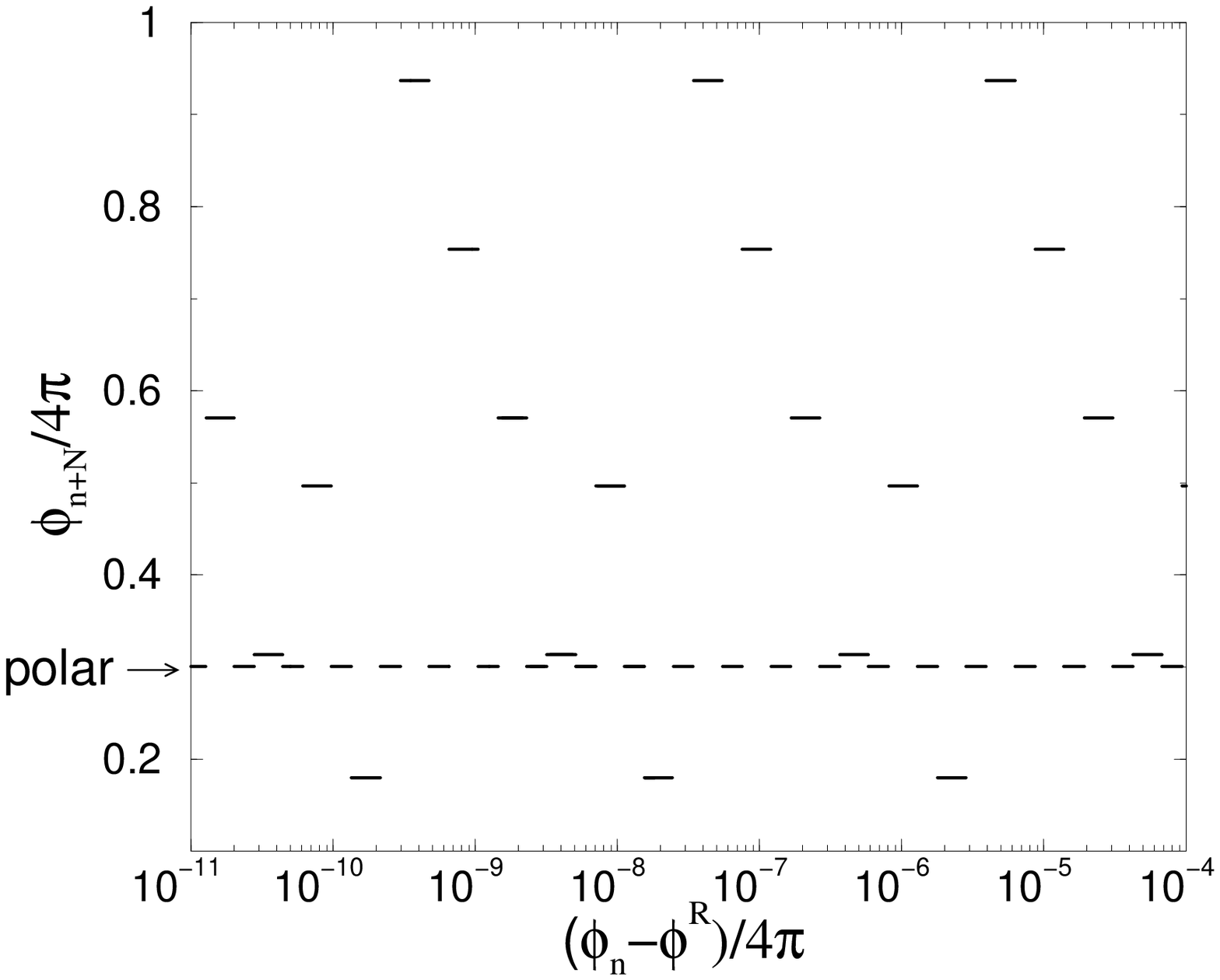}}
\end{minipage}
\caption[]{
Zoom of the $N^{\rm th}$ iterate ($N=1200$) of the map near an
accumulation point. The aspect ratio $\eta$ and the frequency $\omega$
are the same as in figure \ref{fig_map}.  On the left panel we clearly see
the presence of the accumulation point $\phi^R$; the two arrows indicate
the basins of attraction of the polar attractor in figure~\ref{orbs}a;
the other segments belong to the basin of the other (equatorial) attractor.
On the right panel we
see that the width of the intervals of the basins of attraction vanishes
geometrically as the accumulation point is approached.  }
\label{fig_accumul} \end{figure}

When the map \eq{applic} is iterated as in
figure~\ref{fig_map} and since, generically, discontinuities are not mapped
onto themselves\footnote{The case when all discontinuities are mapped onto
themselves corresponds to periodic orbits of the shadow (see
\S\ref{TSC}).}, their number increases
proportionally to the number of iterations; also some fixed points appear
which indicate the existence of attractors, \ie attractive periodic
orbits which we discuss below (\S\ref{sect_attr}).  We would therefore
expect that the basins of attraction of the infinitely iterated map, 
containing an infinite number of intervals at smaller and smaller
scales, would have a fractal structure; however, numerical studies show
only isolated accumulation points which are actually the fixed repulsive
points of the mapping, \ie the repellors (see figure~\ref{fig_accumul}). 

We therefore see that the structure of basins of attraction is much
more complicated than in the case studied by \cite{ML95} and may
represent the general case for such systems.

\subsubsection{Orbits and Lyapunov exponents: the full sphere case}\label{FSC}

Once the mapping is known, we may examine the trajectories of
characteristics.  
For the sake of clarity, it is useful to first consider the case of the
full sphere for which only \eq{cart2} and \eq{cart3} are necessary.

We first note that the number of reflection points of a periodic orbit
is necessarily even, \ie $2q$, when the starting point is not a critical
latitude. Applying alternatively \eq{cart2} and \eq{cart3}, we have

\greq 
\phi_1 = -\phi_2 + 2\lambda \; [2\pi]\\
-\phi_2 = \phi_3 + 2\lambda \; [2\pi]\\
\vdots\qquad\vdots\qquad\vdots\\
\phi_{2q-1} = -\phi_{2q} + 2\lambda \; [2\pi]\\
-\phi_{2q} = \phi_1 + 2\lambda \; [2\pi]
\egreq

\noi Summing all these equations leads to the conclusion that a periodic
orbit is such that

\beq \lambda = \frac{p\pi}{2q} \eeqn{rat_freq}

\noi where $p$ and $2q$ are integers which represent the number of
crossings of the orbit with respectively the axis of rotation or the equator.

From the preceding results, it turns out that all orbits such that
$\lambda = r\pi$ with $r$ irrational, are ergodic (quasi-periodic). At
this point it is worth noting that eigenfrequencies of inertial modes
in the full sphere are in general associated with quasi-periodic
orbits. We prove in appendix~\ref{appA} that, for instance, the first
axisymmetric inertial mode of frequency $\sqrt{3/7}$ is associated with
a quasi-periodic orbit.

To conclude with the full sphere, we just need to point out that thanks
to relations \eq{cart2} or \eq{cart3}, it is clear that the Lyapunov
exponent is always zero. Indeed, if we recall the definition of the Lyapunov
exponent $\Lambda$ associated with an orbit:

\[ \Lambda = \lim_{N\rightarrow\infty} \frac{1}{N}\sum_{n=1}^N
\ln\left|\frac{d\phi_{n+1}}{d\phi_n}\right|, \]

\noi it is clear that for the full sphere
$\left|\frac{d\phi_{n+1}}{d\phi_n}\right|=1\;\forall n$, so $\Lambda=0$.

\subsubsection{Orbits and Lyapunov exponents: the shell case}\label{TSC}

We now turn to the shell case. As we already observed the
map has discontinuities defined by the shadow of the inner
sphere on the outer shell and the projection of critical latitudes.
If the inner sphere is sufficiently small,
the shadow may draw a periodic pattern if the critical latitude $\lambda$ is
commensurable with $\pi$. A simple case is illustrated in
figure~\ref{fig_shad}. For these frequencies, orbits starting in the shadow
remain in the shadow while those starting outside remain outside. In
this way, one can construct the set of all periodic orbits with
$\Lambda=0$ \ie all neutral periodic orbits. 

The fact that periodic orbits outside the shadow are neutral is
obvious; the case of orbits in the shadow is less obvious but we may
consider the fact that if these orbits were not neutral, the shadow would
not map onto itself. As a short exercise, we may follow  one such orbit
for $\omega=\sin (\pi /2(2p+1))$.  The shadow should cross only twice
the location of the inner sphere.  It bounces first on the inner
sphere, at an angle $\beta_1$, and then on the outer sphere at the
angle $\phi_1$ given by $\sin (\phi_1-\delta)= \eta \sin (\beta_1
-\delta )$ ($\delta$ is one of the angles $\delta_n$ in figure~\ref{fig_rec}).
 Then it bounces $2p$  times on the outer sphere. After
each two reflections, the angle $\phi$ changes into $\phi+ 4 \delta $.
Therefore after $2p$ bounces, the angle on the outer sphere is
$\phi_1 + 4 p \delta$.  It bounces then again on the inner sphere,
hitting it at the angle $\beta_2$ such that $\sin (\phi_1+\delta +
4 p \delta)= \eta \sin (\beta_2 +\delta )$.  But $2(2p+1) \delta
=\pi$ therefore $\delta +  4 p \delta = \pi - \delta [2\pi]$.  So
$\sin (\phi_1-\delta )= -\eta \sin (\beta_2 +\delta )= \eta \sin
(\beta_1 -\delta )$.  Therefore $\beta_1 =-\beta_2 [2\pi]$ or $\beta_1
=\pi +\beta_2 +2\delta [2\pi]$.  Repeating this entire process
$2(2p+1)$ times, one comes back to the original $\beta_1$.

In fact periodic orbits of the shadow do not exist for all $\lambda$'s
commensurable with $\pi$. Indeed the image of the shadow must not be
split by discontinuities; this implies that $p$ or $q$ cannot be too
large or the shell too thin. More precisely, for a given $\eta$,
periodic orbits of the shadow exist if:

\beq \arcsin\eta \leq \lambda \leq \arccos\eta \eeqn{freq_shad}

If $\eta\geq 1/\sqrt{2}$, only one
neutral periodic orbit exists: it is such that $\lambda=\pi/4$ or
\mbox{$\omega=1/\sqrt{2}$.} More generally, for a given aspect ratio $\eta$,
one may determine all the frequencies associated with neutral periodic
orbits and their number increases as the size of the inner shell
decreases.  These frequencies are obviously determined  by \eq{rat_freq} but due
to the finite size of the shadow, one needs to eliminate large values of
$p$ and $q$. If we remark that the most robust periodic orbits (when
$\eta$ increases) are those with
$p=1$ (restricting ourselves to $\lambda\leq\pi/4$), relation
\eq{freq_shad} easily bounds the permitted values of $q$. For a shell
like the core of the Earth, where $\eta=0.35$, only $q=2,3,4$ are
possible.

\begin{figure}
\centerline{\includegraphics[width=7cm,angle=90]{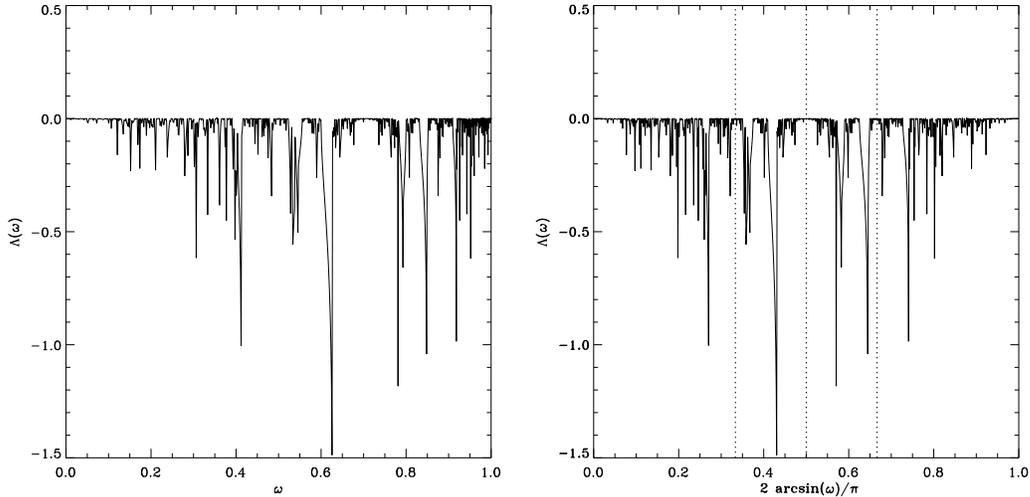}}
\caption[]{The lyapunov exponent of one basin as a function of
frequency (left) or critical latitude (right). Note the symmetry of
this last plot with respect to $\pi/4$. The vertical dotted lines mark
the critical latitudes $\pi/6$, $\pi/4$ and $\pi/3$; the aspect ratio
of the shell is $\eta=0.35$.} \label{lyap} \end{figure}

The frequencies of neutral periodic orbits
are important as they shape the form of the Lyapunov
exponent curve since, then, intervals contracting through $f_+$ are
exactly dilated by $f_-$ which makes the Lyapunov exponent vanish.
As a consequence, frequencies in the neighbourhood of one such frequency
are associated with very long orbits having a small Lyapunov exponent.
This is the reason why `windows' appear near these frequencies,
especially near $\omega=1/\sqrt{2}$ ($\lambda=\pi/4$) as clearly shown in
figure \ref{lyap}.

This latter figure shows many spikes which in fact betray the
presence of other periodic orbits which we shall call {\em attractors} after
\cite{ML95}. For such orbits $\Lambda\leq 0$. In fact for this system,
all the orbits (except may be some isolated ones) verify this
inequality and no chaos is possible: the configuration space being
one-dimensional and the mapping being one-to-one.

\subsubsection{Some properties of attractors in the shell}\label{sect_attr}

\begin{figure}
\centerline{\includegraphics[width=7cm,angle=-0]{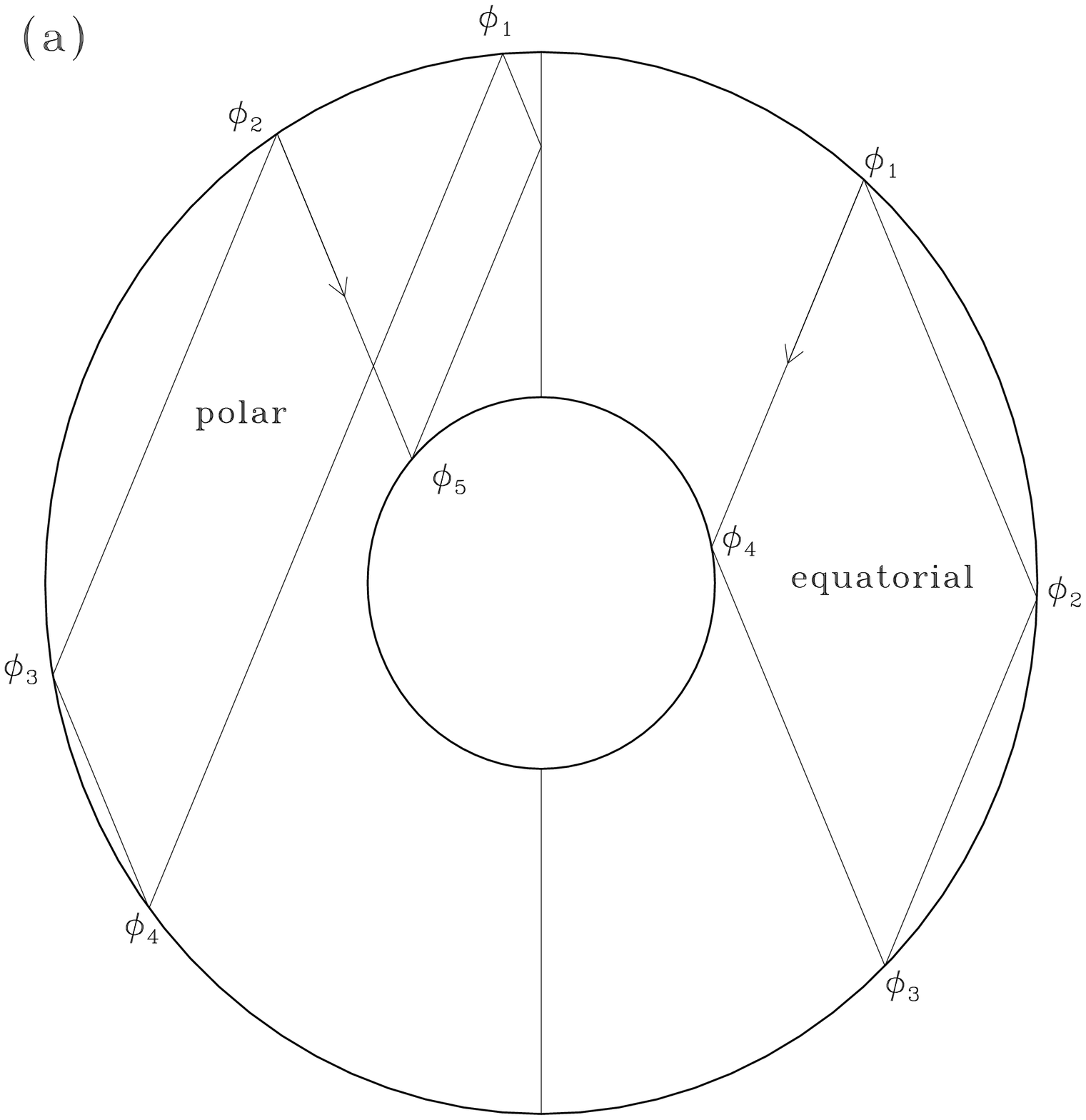}
\includegraphics[width=7cm,angle=-0]{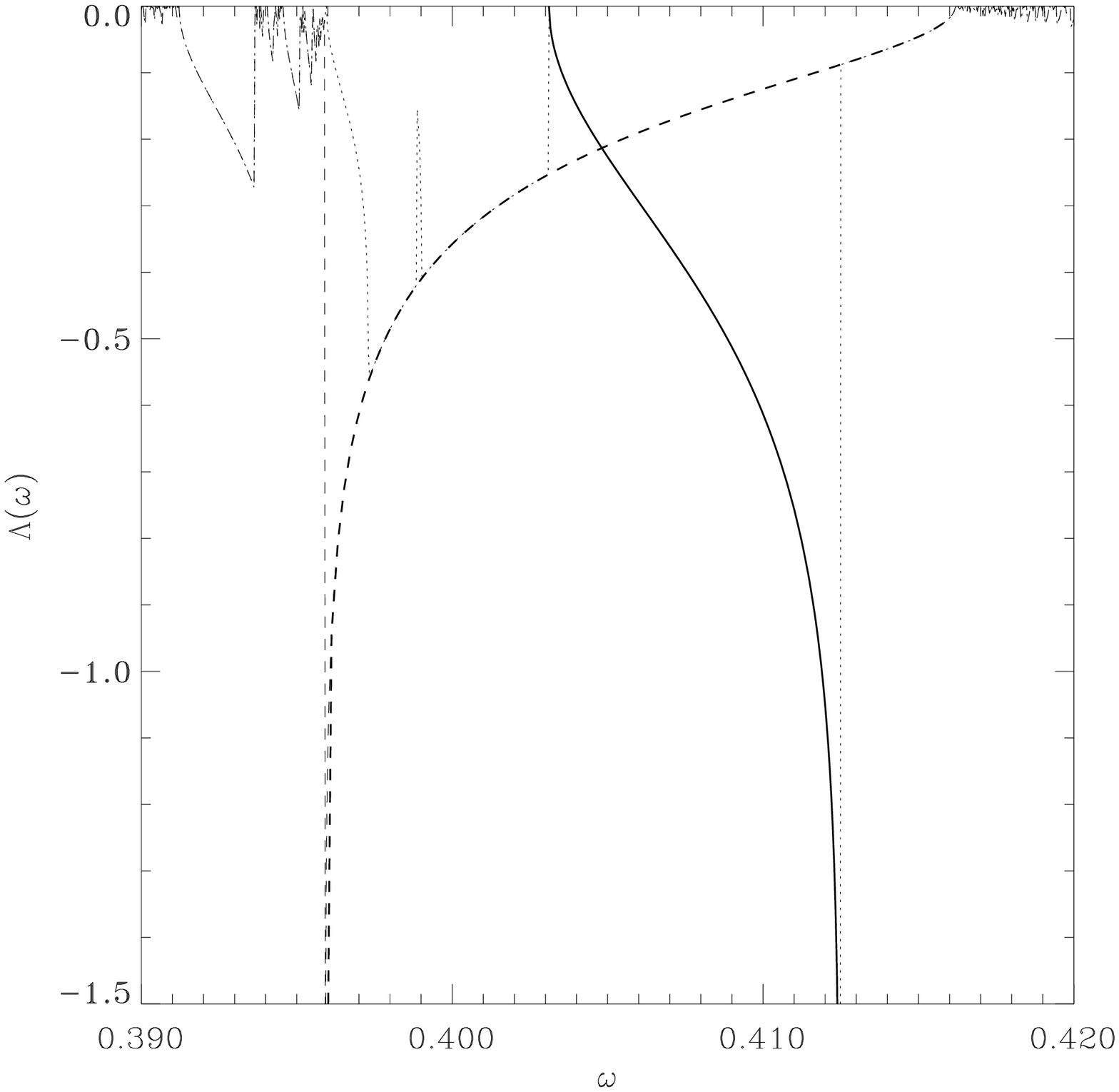} }
\caption[]{(a) Two attractors coexisting when $\omega=0.4051$; we call
them `polar' and `equatorial' respectively as they are
characterized by the fact that the reflection on the inner sphere occurs
above or below the critical latitude; the arrows indicate the direction
of focusing. (b) Lyapunov exponent
as a function of frequency in the vicinity of $\omega=0.4051$; the thick
solid line denotes the equatorial attractor while the thick dashed line
represents the polar attractor; these two thick lines have been derived
from the analytic formulae \eq{lyap_eq405} and \eq{lyap_pol405} given
in appendix~\ref{appC}.}
\label{orbs}
\end{figure}

In order to make the dynamics of attractors
clearer, it is useful to concentrate on a specific example
which can be computed explicitly. For this purpose, we choose a
spherical shell similar to that of the core of the Earth for which
$\eta=0.35$ and we focus on the orbit with four reflections on the
outer or inner shell as shown in figure \ref{orbs}. If the shell is
thinner, this orbit is localized in the vicinity of the equator which is
the kind studied by \cite{stewar71,stewar72a,stewar72b}.

The Lyapunov exponent of this orbit may be computed explicitly in the
following way: Let us first recall that plane inertial waves reflecting
on a surface oriented by $\vn$ verifies the relation

\[ \vk_i\times\vn = \vk_r\times\vn \]

\noi where $\vk_i$ and $\vk_r$ are the wave vectors of the incident and
reflected waves respectively \cite[]{Green69}. When applied to a
reflection on a sphere, this relation implies

\[ k_r = k_i\frac{\sin(\phi\pm\lambda)}{\sin(\phi\mp\lambda)} \]

\noi where $\phi$ is the `latitude' of the reflection point. From this
relation, we can compute the rate of contraction of an infinitesimal
interval in the neighbourhood of a reflection point of the orbit.
Therefore, the Lyapunov exponent of an attractor with $N$ reflection
points is simply given by:

\beq \Lambda = -\frac{1}{N} \ln \prod_{k=1}^N
\frac{\sin(\phi_k\pm\lambda)}{\sin(\phi_k\mp\lambda)} \eeqn{Lyap_form}

\noi where the $\phi_k$'s describe the reflection points of the
attractor (a periodic orbit). At this point we shall define two useful
quantities characterizing attractors, namely their {\em length} and their {\em
period}. We shall call the number of reflection points (on both spheres)
the length of the attractor while its period will refer to the minimum
number of iteration of the mapping needed to generate its associated
fixed points; because of the definition of the mapping, the period is also
the number of reflection points on the outer sphere. For instance, the
period of the equatorial attractor of figure~\ref{orbs}a is '3' and its
length is '4', while for the polar one, the length is '10' and the
period '8' (we do not allow reflections on the axis).
We give in appendix~\ref{appC} the explicit
calculations relevant to the orbits of figure \ref{orbs}.

The curves $\Lambda(\omega)$ (figure \ref{orbs}b)
show the same feature: in the interval of
existence of the orbit, $\Lambda$ varies between 0 and $-\infty$. The two
extremes correspond respectively to the cases when the reflection on the
inner shell occurs at the equator or at its critical latitude. We note
that the vanishing value of the Lyapunov exponent does not mean that for
this frequency the orbit is no longer an attractor: it simply
means that convergence towards the attractor is no longer exponential;
in most cases, it does converge, but algebraically.

For later use, we also computed the behaviour of $\Lambda(\omega)$ in
the vicinity of the point $\omega_0$ such that $\Lambda(\omega_0)=0$. We
find in the particular case of the equatorial attractor that

\beq \Lambda(\omega) = -K(\omega-\omega_0)^{1/2}, \qquad {\rm with}\quad
K=4.8184, \quad \omega_0=0.403112887 \eeqn{as_lyap}

\noi In fact this behaviour is general as is shown in appendix~\ref{appC}.
Let us also emphasize that when
$\Lambda=0$, the orbit is just one broken line connecting:
\begin{itemize}
\item the equator or the pole of the spheres to another equator or pole,
\item the equator or the pole of the spheres to the critical latitudes of
the outer sphere.
\end{itemize}

\noi With the terminology in use for dynamical systems,
such an orbit restricted to the first quadrant is self-retracing: a
mass-point would go back and forth on the same trajectory.

The foregoing results make the shape of figure \ref{lyap} quite clear now.
Most of the `spikes' shown in this graph will therefore tend to $-\infty$ as
the number of points of the graph is increased. But to be complete,
we need also mention some cases when a segment of an orbit intercepts the
inner shell after being tangential to it. In this case the curve
$\Lambda(\omega)$ has a discontinuity and does not reach $-\infty$.

We therefore see that attractors are featuring figure \ref{lyap}. As it
will be clear later they also feature the shape of the asymptotic
spectrum. It is thus interesting to know some elementary properties of
these geometrical objects.

From a rather large number of computations we observed, as \cite{ML95},
that, in the first quadrant, not more than two attractors may coexist
for a single frequency.  However, these two attractors can be used to
construct other attractors which are just their image symmetrized
with respect to axis of rotation and equator. Considering
the propagation of characteristics in the full meridional section of
the shell, we observe that these lines can converge towards six
attractors at most. Using the properties of the mapping~\eq{applic}, we
have been able to prove under certain hypotheses (see
appendix~\ref{appB}), that the number  of attractors is bounded by the
number of points of discontinuity (which is twelve).

\begin{figure}
\hfil
\begin{minipage}[t]{.45\linewidth}
 \centerline{\includegraphics[width=\linewidth]{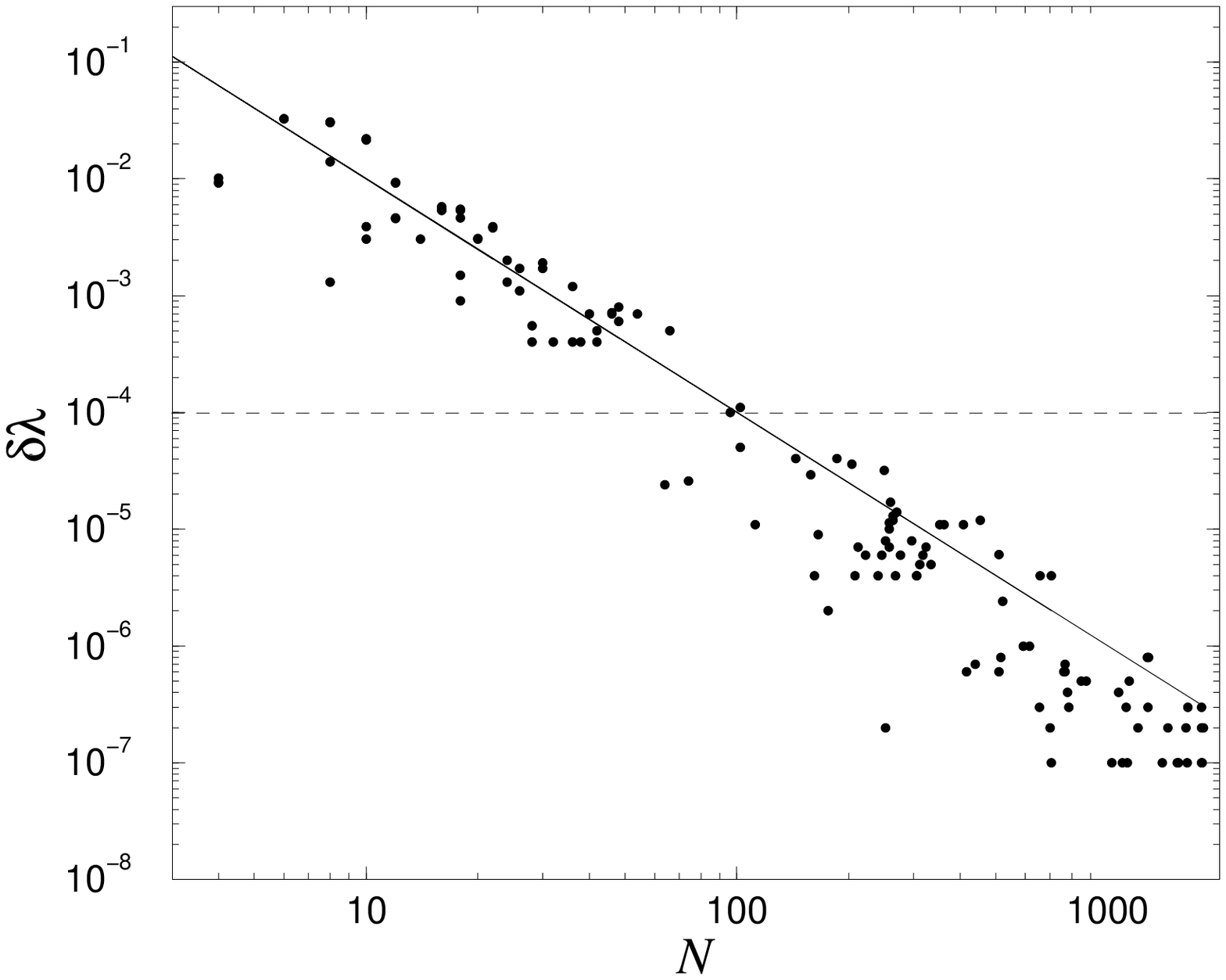}}
\caption[]{Interval of existence in latitude of several different
attractors, plotted as a function of their length $N$
The straight solid line is the inverse square of the length.}
\label{period_freq}
\end{minipage}
\hfil
\begin{minipage}[t]{.45\linewidth}
 \centerline{\includegraphics[width=\linewidth]{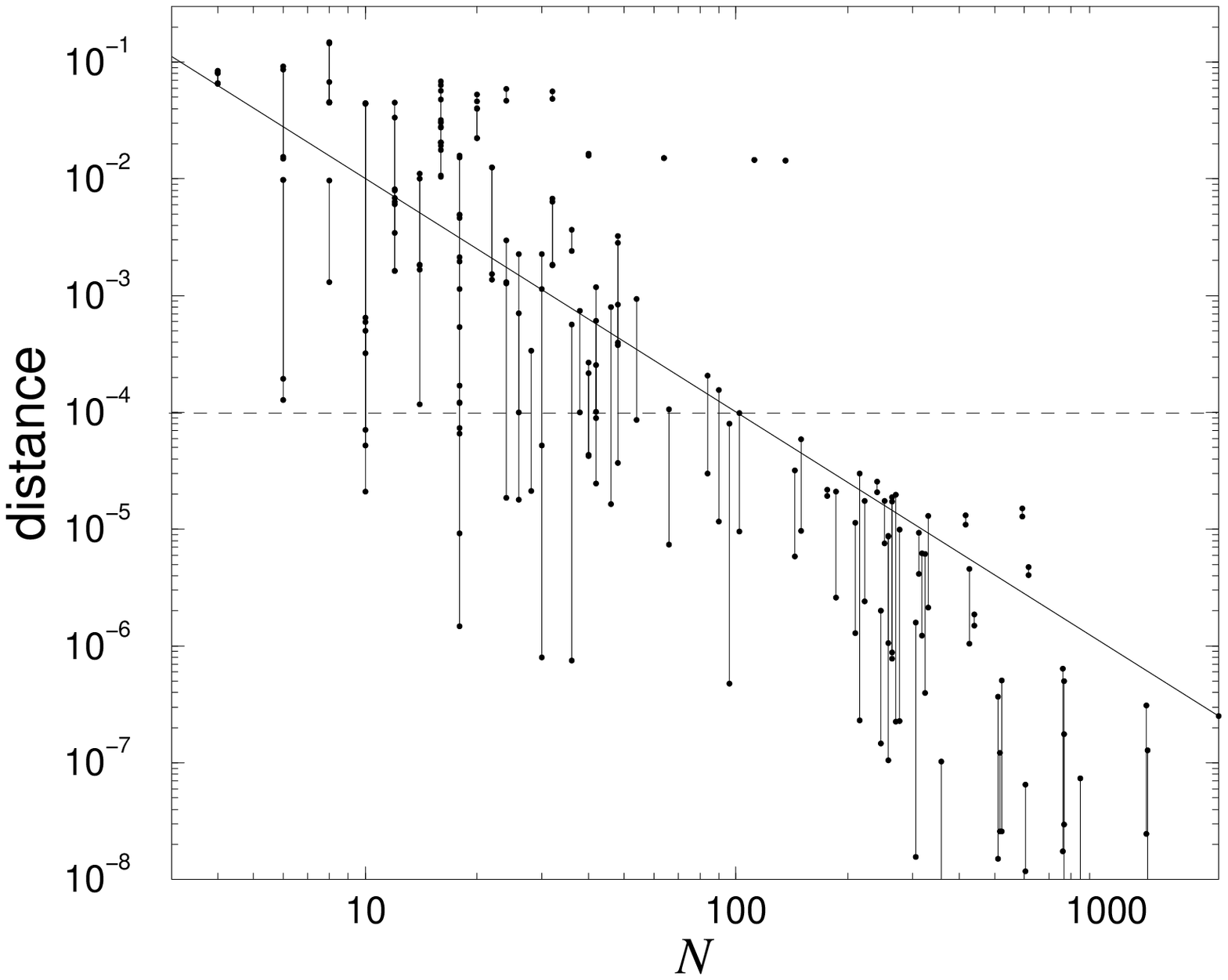}}
\caption[]{Distance 
to the point on the external sphere at critical latitude,
for several different attractors, plotted
as a function of the length $N$.
The straight solid line is the inverse square of the length.
The dashed line gives the lower bound (in distance) for physically
relevant attractors (see discussion in \S\ref{thex}).
}\label{period_distance}
\end{minipage}
\hfil
\end{figure}

We also computed the interval of existence, in frequency space,
of a large number of attractors so as to show its relation with the
length of the attractor.
As shown in
figure~\ref{period_freq}, the interval of existence is well correlated
with the inverse square of the length. We explain this law in the
following way: for a very long attractor of length $N\gg1$ the number of
reflections on the inner and outer shell scales with $N$, therefore the
mean angular distance between the critical latitude and the nearest
reflection point is \od{1/N}. This implies that just a \od{1/N^2}
variation in frequency is necessary to shift this point to the critical
latitude.

The latter result has an interesting consequence on the shape of the
curve $\Lambda(\omega)$ for a given long attractor. Indeed,
from~\eq{Lyap_form} the divergence toward $-\infty$ of an attractor of
length $N$ is of the form:

\[ \Lambda \sim \frac{1}{N}\ln\lc N(\omega-\omega_c)\rc \]
where $\omega_c$ is the frequency of the singularity of $\Lambda$;
we used the fact that \eq{Lyap_form} is dominated by one term
( $\phi_k-\lambda\sim 0$ with $\phi_k-\lambda\sim N(\omega-\omega_c)$ );
since the interval of existence of the attractor scales like $1/N^2$, if
we choose a point such that $\omega-\omega_c= \alpha/N^2$ 
the Lyapunov exponent will scale like $\Lambda \sim
-{\ln(\alpha N) \over N}$ which vanishes at large $N$. This means that
the singularity of the Lypaunov curve occupies a smaller and smaller
fraction of the interval of existence of the attractor. Hence for long
attractors, the Lyapunov exponent is very small in a larger and larger
part of their interval of existence. This explains why long attractors
appear numerically as weakly attracting eventhough their Lyapunov
exponent may diverge.

In figure~\ref{period_distance} we show the distance of attractors to
the point on the external sphere at critical latitude as a function of
the length.  Since a periodic orbit exists in a range of frequencies
(see figure~\ref{period_freq}), instead of showing a single point we
represent a vertical segment connecting the minimum and maximum
distance over the entire range of existence of the attractor in
frequency space.  We see that the maximum distance is well correlated
with the inverse square of the length.  This distance is important in
the final appearance of the attractor when viscosity is included. As it
will be shown later on an example, the build-up of energy along an
attractor can be impeded by the boundaries; this effect therefore puts
an upper bound on the viscosity for the attractor to be visible. The
dashed line in the figure gives the lower bound (in distance) for
physically relevant attractors (see discussion and the example in
\S\ref{thex}).

\subsubsection{Differences with billiards}

It is interesting to compare the mapping defined by
(\ref{cart1}-\ref{cart3}) with the billiard problem studied in
classical chaos, where a particle bounces specularly on the walls of a
cavity.  Billiard phase spaces are two-dimensional (position and
velocity direction) while the phase space of our problem is
one-dimensional (in projection), since the only variable is the
position along the circles representing inner and outer shells.  The
problem is not Hamiltonian, there is no conservation of the symplectic
measure in phase space, and attractors and repellors exist.

For the full sphere, $\eta=0$, we have seen that all the orbits are
neutral ($\Lambda=0$) and are either quasiperiodic (and ergodic) or
periodic (when $\lambda$ is a rational fraction of $\pi$).  When $\eta$
is made nonzero, all quasiperiodic orbits are instantaneously
destroyed, and the periodic orbits remain neutral until they are
eventually destroyed when $\eta$ increases, the rational values with
smaller denominator surviving last.  It is interesting to note that
this situation is exactly the opposite of the Kolmogorov-Arnold-Moser
(KAM) theorem valid for Hamiltonian systems close to integrability
\cite[see][]{arnold89}.  In this latter case, it is well-known that
for an integrable
Hamiltonian system all orbits lie on tori, and orbits are organized in
families, either quasiperiodic (and ergodic) or periodic, for a given
torus.  If one perturbs such an integrable Hamiltonian system by a
sufficiently smooth perturbation, the KAM theorem states first that all
rational tori (with periodic orbits) are instantaneously destroyed, and
second that irrational tori (with quasiperiodic orbits) disappear one
after the other when the perturbation is increased, the last to
disappear being the `furthest' to the rationals.

\subsection{Relations between orbits of characteristics and eigenfunctions}
\label{stream}

In the preceding subsections we have shown that, generically,
characteristics converge towards attractors which are formed by a
periodic orbit. These attractors live in a frequency band whose
size decreases with the length (or period) of the attractor. The
attracting power is measured by a negative Lyapunov exponent which
generically varies between 0 and $-\infty$ when the frequency band is
scanned. Several (less than 6) attractors may coexist for a given
frequency; in this case they own each a basin of attraction (described
for instance as a set of points on the outer boundary) whose structure
is governed by accumulation points (see fig.~\ref{fig_accumul}).

We have also found periodic orbits which are {\em not} attractors; their
frequency can be written $\omega=\sin(p\pi/2q)$ where $(p,q)$ are chosen
in a finite set of integers. The number of these orbits is therefore finite
but increases as the radius of the inner core decreases. The
frequencies of these orbits will prove to be useful since in their
neighbourhood, attractors have very great length and, therefore, very
small (in absolute value) Lyapunov exponents. This will influence the
shape of the asymptotic spectrum (\ie with low viscosities).

Now we shall see how the eigenfunctions are influenced by the presence
of an attractor.

\subsubsection{The two-dimensional case}

Because of the simple form of the Poincar\'e equation in two dimensions, which
may be written as

\beq \dupdum{P}=0 \eeqn{poinc2d}

\noi early
investigations have focused on this case in particular those of
mathematicians. The relevant contributions are those of
\cite{BD39}, \cite{John41}, \cite{hoil62}, \cite{Frankl72}, \cite{Rals73}
and \cite{Schaef75}. Much of this
previous work is concentrated in a theorem demonstrated in
\cite{Schaef75} which states that:

{\it There are non-trivial solutions of \eq{poinc2d} if and only if there
exists an integer n such that all reflected rays close after precisely
2n reflections. If there is one solution, then there are infinitely
many, linearly independent solutions.}

In other words, eigenvalues of regular modes
are always associated with periodic orbits
and these eigenvalues are always infinitely degenerate. Since the
reflected rays must close, starting from any point of the curve, the
Lyapunov exponent of such periodic orbits is always zero. Note the
difference with the three-dimensional case where eigenmodes in the full
sphere are associated with ergodic (quasi-periodic) orbits (cf
\S\ref{FSC} and appendix~\ref{appA}) and eigenvalues are non-degenerate.

Another interesting result was derived from mathematical analysis by
\cite{Rals73}. Namely, it states that the velocity field associated with
a solution of \eq{poinc2d} is not square-integrable when characteristics are
focused towards a wedge formed by the boundaries.
An example of such a singular flow is given in \cite{W68} with the case
of internal waves focused by a sloping boundary. In the interval of
frequencies where the velocity field is not square-integrable,
eigenvalues do not exist and the point spectrum of Poincar\'e operator
is said to be empty.

The foregoing results may be generalized to our case, or the one studied by
\cite{ML95}, where characteristics are focused towards an attractor.
Indeed, let us consider the total kinetic energy of a `mode' associated
with an attractor; using characteristic coordinates, this quantity reads

\[ I = \int_S \|\vv\|^2du_+du_- + \int_{CS} \|\vv\|^2du_+du_- \]

\noi where $S$ designates a neighbourhood of the attractor and $CS$ the
remaining `volume'; we assume that the limits of $S$ are made up of
characteristics. If the attractor is of length $N$ then this integral
can be split into $N$ pieces 

\[ I = \sum_{n=1}^N \int_{S_n}  \|\vv\|^2du_+du_- \]

\noi where we neglected the contribution from $CS$.
Now each of these pieces can be split again into an infinite number
of rectangles $R_k$ with sides made up of characteristics. Hence we write

\[ \int_{S_n}  \|\vv\|^2du_+du_- = \sum_{k=1}^\infty
\int_{R_k}\|\vv\|^2du_+du_- = \sum_{k=1}^\infty I_k\]

\noi In the vicinity of the
attractor, these rectangles are very elongated: one side remains \od{1} long
while the other shrinks to zero as the attractor is approached. 

Now we take the two long sides as images through the mapping made by
the characteristics. The mapping has a contracting rate given by
$e^{\Lambda} <1$ where $\Lambda$ is its Lyapunov exponent. \cite{ML95}
have shown how one can construct the stream function in the whole domain
by iterating an arbitrary function given on its boundary. When the
attractor is approached, the scale of the stream function vanishes while
its amplitude remains constant; therefore the kinetic energy is
amplified by a factor $e^{-2\Lambda}$ at each iteration of the mapping.
Noting that one rectangle is smaller by a factor $e^{\Lambda}$ than its
predecessor, we can derive the iteration rule 

\[ I_{k+1} = e^{-\Lambda} I_k \]

\noi which shows that the integral $I$ is infinite. We may note in
passing that if $d_k$ is the distance of the $k^{\rm th}$
characteristic to the limit cycle, then $d_k = d_0 e^{k\Lambda}$ while
the amplitude of the velocity field is $v_k = v_0 e^{-k\Lambda}$. This
shows that the velocity field diverges as the inverse of the distance to
the attractor.

Therefore, as in the case of a wedge, the velocity field is not 
square-integrable when characteristics converge towards an attractor.

\subsubsection{The three-dimensional case}

The 3-D case has not benefitted from the same interest by
mathematicians. In this case the Poincar\'e equation contains first or
zeroth order derivatives which cannot be eliminated. Let us rewrite it
using cylindrical coordinates and assume a exp$(im\varphi)$ dependence
of the pressure; thus

\[ \dds{P} +\frac{1}{s}\ds{P} -\frac{\alpha^2}{\omega^2}\ddz{P} -
\frac{m^2P}{s^2} =0 \]

\noi The canonical form of this equation is obtained using characteristics
coordinates:

\beq 2\dupdum{P}
-\frac{1}{u_+-u_-}\lp\dup{P}-\dum{P}\rp - \frac{m^2}{(u_+-u_-)^2}P =0
\eeqn{poinc3d}

\noi which is known as the Euler-Darboux equation \cite[]{DL84}.

\psfrag{M}{S}\psfrag{P}{P}\psfrag{Q}{Q}
\psfrag{S}{\hspace*{1mm}\raisebox{-0.4mm}{M}}
\psfrag{X}{\raisebox{0.2mm}{$\bullet$}}
\psfrag{s}{$s$}\psfrag{z}{$z$}
\begin{figure}
\centerline{\includegraphics[width=9.5cm,angle=0]{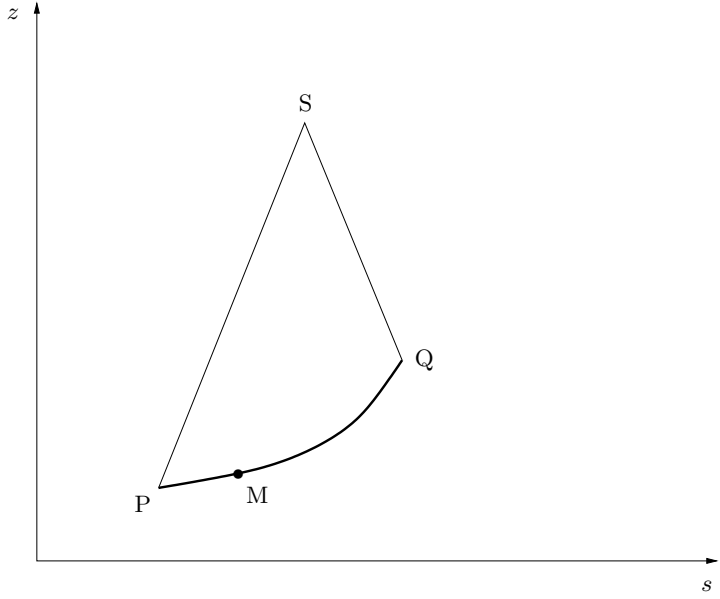}}
\caption[]{A sketch for the illustration of Riemann's method; the lines
(SP) and (SQ) are segments of characteristics.}
\label{rieman}
\end{figure}

A general solution of Euler-Darboux equation may be obtained with
Riemann's method \cite[]{colombo76,Zwilling92}.
With this method one may express the value of the solution at one 
point when `initial' data are given on
an arc joining two points on characteristics `emitted' from the point
considered (see figure~\ref{rieman}). However, one needs to know the 
Riemann function
(which plays an equivalent role to the Green's function of elliptic
problems). To determine this function it is useful to rewrite the
pressure fluctuation as $P=\Pi/\sqrt{s}$; doing so, the first derivatives of
Poincar\'e equation are eliminated but are replaced by the term $\Pi/4s^2$.
The equation for $\Pi$ is therefore:

\beq \dupdum{\Pi} + \lp m^2-\frac{1}{4}\rp\frac{\Pi}{(u_+-u_-)^2} =0
\eeqn{poinc_new}

The Riemann function is a solution of this equation\footnote{In fact,
it is a solution of
the corresponding adjoint operator which is identical to the original in
this case.} which meets the additional conditions:

\beq R(u_+,u_-)=1, \qquad \left.\dup{R}\rp_{u_-}=\left.\dum{R}\rp_{u_+}=0 \eeq

\noi or, equivalently,

\[ R(u'_+,u_-)=R(u_+,u'_-)=1 \qquad \forall (u'_+,u'_-)\in {\cal D} \]

\noi We shall call $S(u_+,u_-)$ the point where the solution is computed
and $M(u'_+,u'_-)$ a point running on the arc of data; ${\cal D}$ is
the area defined by (SPQ).

As \cite{FH68} have noted, \eq{poinc_new} is invariant for all
the transformations leaving

\[ z = \frac{(u'_+-u_+)(u'_- -u_-)}{(u'_+ - u'_-)(u_+-u_-)} \]

\noi invariant. Therefore, seeking a solution of the form $R(z)$,
one finds that this function verifies the differential equation:

\[ z(1-z)R'' - (2z-1)R' + \mu(\mu-1) R =0 \]

\noi where we set $\mu=m+1/2$. This is a special case of the
differential equation of Gauss hypergeometric function, \ie
$F(\mu,1-\mu;1;z)$; in fact, it is just the equation satisfied by Legendre
functions of index $\mu-1$.

Since $z\equiv z(S,M)$, we shall write Riemann's function as $R(S;M)$.
Hence the formal solution of the problem is

\beqan &&\hspace*{-1.5cm}\lefteqn{\Pi(S)= \demi(\Pi(P)+\Pi(Q))}  \nonumber\\
&&\hspace*{-1cm}+\demi\int_{PQ} R(S;M)\lp
\dup{\Pi}du_+ - \dum{\Pi}du_-\rp + \Pi(M)\lp\dup{R}du_+ -
\dum{R}du_-\rp \eeqan{gen_sol}
which shows how the value of $\Pi$ at $S$ may be constructed from
the data given on the arc $PQ$.

A simpler formula can be obtained for axisymmetric modes when the
meridional stream function $\psi$ is considered\footnote{This function
is such that 
$ u_s = \frac{1}{s}\dz{\psi}, \quad u_z = - \frac{1}{s}\ds{\psi} $
}.  After a similar transformation, where we set $\psi=\sqrt{s}\Psi$, we
find that the associated Riemann function is also a Legendre function
with $\mu=-1/2$. If the arc $PQ$ is taken on the boundary then $\Psi=0$
and the expression \eq{gen_sol} simplifies into

\beq \Psi(S)=\demi\int_{PQ} R(S;M)\lp \dup{\Psi}du_+ - \dum{\Psi}du_-\rp
\eeq

Let us now suppose that $S$ is also on the boundary (on the inner sphere
for instance); then $\Psi(S)=0$. If we introduce $\dcp=\lp
\dup{\Psi}du_+ - \dum{\Psi}du_-\rp$, then we have

\[ \int_{PQ} R(S;M)\dcp = 0 \]

This relation holds also for neighbouring points $(P'Q'S')$ of $(PQS)$, thus

\[ \int_{P'Q'} R(S';M)\dcp = 0 \]

\noi By subtracting these two equations we get

\beq \int_{PP'}R(S;M)\dcp + \int_{QQ'} R(S;M)\dcp + d\phi_S\int_{PQ}
\dphi{R}(S;M)\dcp = 0 \eeqn{drel}

\noi where $d\phi_S$ denotes the variation of the position of $S$.
Since $P'$ and $Q'$ are in a neighbourhood of $P$ and $Q$
respectively and that $R(S;P)=R(S;Q)=1$, the first integrals of
\eq{drel} can be simplified so that:

\beq {\cal C}(\Psi)(P) + {\cal C}(\Psi)(Q) +
\dnphi{\phi_S}\int_{PQ}\dphi{R}(S;M)\dcp = 0 \eeqn{drel2}

\noi where ${\cal C}(\Psi)=\dup{\Psi}\dphi{u_+}-\dum{\Psi}\dphi{u_-}$.

Now we consider that $PSQ$ are part of a
limit cycle like the one of figure~\ref{orbs}a (right); let us call
$T$ the fourth
point of this cycle and let $P_n,S_n,Q_n,T_n$ be the suite of points
converging towards $PSQT$ (\ie those points at $(\phi_3, \phi_4,
\phi_1, \phi_2)$ respectively). \eq{drel2} can be applied to the triangles
$P_n,S_n,Q_n$ and $Q_n,T_n,P_{n+1}$ and we get:

\[ {\cal C}(\Psi)(P_n) = {\cal C}(\Psi)(P_{n+1}) +
\dnphi{\phi_T}\int_{Q_nP_{n+1}}\dphi{R}(T_n,M)\dcp +
\dnphi{\phi_S}\int_{Q_nP_n}\dphi{R}(S_n,M)\dcp \]

The two integrals in the RHS are of order unity and we surmise that
they do not cancel.  Therefore the suite of ${\cal C}(\Psi)(P_n)$ is
diverging which means that the velocity fields tends to infinity when a
limit cycle is approached.

This result can be generalized to any limit cycle. We may also use it to
generalize Ralston's theorem. However, in this latter case, it is more
straightforward to note that if we are considering characteristics
converging towards a wedge (for a three-dimensional problem), then in a
neighbourhood of the apex of the wedge, first order derivatives are
negligible compared to second order derivatives; therefore, in this
neighbourhood, \eq{poinc3d} can be transformed into \eq{poinc2d} and
Ralston's theorem applies.

\subsection{The critical latitude singularity}

The preceding discussion has shown (and proved in 2D) that the velocity
field of `modes'\footnote{We use quotes because modes refer usually to
regular solutions with square-integrable velocity fields.} associated
with an attractor is not square-integrable: it diverges
as the inverse of the distance to the attractor.

We shall see now that this singularity is not the only one and that a
milder one develops around the critical latitude of the inner sphere.
\cite{StRic69} were the first to notice that singularity and showed
that it is integrable. Although the work of \cite{StRic69} was restricted
to the thin shell limit and was based on the use of Longuet-Higgins solutions
of the Laplace tidal equation, we shall show that their result is
in fact general.

For quick reading demonstrations in \S\ref{sscl} and \S\ref{vfcl} can be
skipped.

\subsubsection{The singular surfaces}\label{sscl}

Let us consider a sphere immersed in a rotating fluid
filling the whole space. We examine the oscillations of the fluid. Such
modes are the corresponding modes of the full sphere when solutions
regular at the origin are replaced by solutions regular at infinity.

As for the full sphere, we use Bryan's transformation to convert
the Poincar\'e equation into the Laplace equation. Thus we set

\[ z'=-i\frac{\omega}{\alpha} z \]

To solve Laplace's equation, we therefore use ellipsoidal
coordinates $(\xi,\theta,\varphi)$ similar to the spherical coordinates
(for the angular variables $\theta$ and $\varphi$):

\greq
x = a\cosh\xi\sin\theta\cos\varphi \\
y = a\cosh\xi\sin\theta\sin\varphi \\
z' = a\sinh\xi\cos\theta
\egreq

\noi where $a$ is the focal distance of the meridional ellipse.
If we take the radius of the sphere as the unit of length, then
$a=1/\alpha$. We recall that using these coordinates, Laplace equation
of an axisymmetric field transforms into :

\beq \lap V =
\frac{1}{\cosh\xi}\dxi{}\lp\cosh\xi\dxi{V}\rp + \dsint{V} =0 \eeq

\noi whose solution, regular at infinity, reads:

\beq Q_\ell(i\sinh\xi)P_\ell(\cth) \eeq

\noi where $Q_\ell$ is the second-kind Legendre function.

If we now come back to the original coordinates, we may write the cylindrical
coordinates $(s,z)$ as

\greq
\disp{s = \frac{1}{\alpha}\cosh\xi\sth}\\
\\
\disp{z = \frac{i}{\omega}\sinh\xi\cth}
\egreqn{sys1}

\noi and following the idea of \cite{Green69}, we introduce

\beq \mu = \cth \qquad {\rm and}\qquad \eta = i\sinh\xi \eeq

\noi so that

\greq
\alpha s = \sqrt{(1-\mu^2)(1-\eta^2)} \\
\omega z = \mu\eta
\egreqn{trsf2}

\noi The Jacobian of this new coordinate transform is 

\beq J = \frac{\alpha^2\omega s}{\eta^2-\mu^2} \eeq

\noi where we used

\greq
\ds{\mu} = \frac{\alpha^2s\mu}{\Delta} , \qquad \dz{\mu} =
\frac{\omega(1-\mu^2)\eta}{\Delta} \\
\\
\ds{\eta} = -\frac{\alpha^2s\eta}{\Delta} , \qquad \dz{\eta} =
-\frac{\omega\mu(1-\eta^2)}{\Delta}
\egreqn{transf_coo2}

\noi with $\Delta = \eta^2-\mu^2$. Hence, the transform is singular
on the surfaces such that $\eta = \pm\mu$. Since the solution
$Q_\ell(\eta)P_\ell(\mu)$ is regular in the fluid's domain, the
singularity of the transformation makes the solutions singular when the
coordinates map the space in a regular way.

To discover which kind of surfaces hinds behind this equation ($\eta =
\pm\mu$), it is convenient to express $\xi$ (or $\eta$) as a function of the
cylindrical coordinates $s$ and $z$. Eliminating $\theta$ from \eq{sys1}
and setting $X=\sinh^2\xi$, we find that 

\[ X^2 + (1+\omega^2z^2-\alpha^2s^2)X + \omega^2z^2 = 0 \]

The solution of this equation gives the reciprocal transformation of
coordinates \eq{trsf2} or \eq{sys1}. When the roots are multiple,
the transformation is singular; this happens when the discriminant $D$
vanishes which is when

\beq D=(1-\omega z -\alpha s)(1-\omega z +\alpha s)(1+\omega z
+\alpha s)(1+\omega z-\alpha s) = 0 \eeq

\noi One may easily verify that this equation is equivalent to
$\eta^2=\mu^2$.

\begin{figure}
\centerline{\includegraphics[width=8cm,angle=-90]{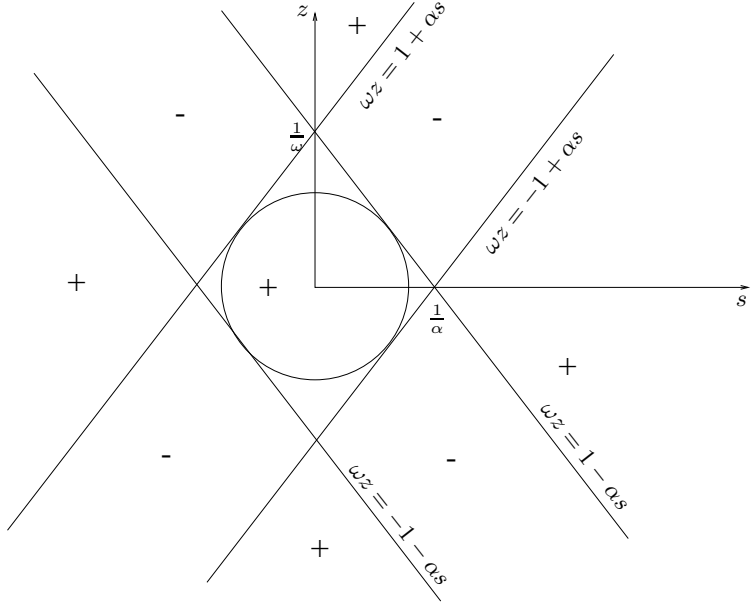}}
\caption[]{Meridional cross-section of the surfaces (cones) where the
coordinate transform is singular. The signs $+/-$ refer to the sign of the
discriminant $D$.}
\label{discri}
\end{figure}

We have therefore shown that the transformation is singular on four surfaces
which are cones tangent to the sphere at the critical latitudes.
This result is summarized in figure~\ref{discri}.

\psfrag{d1}{\begin{rotate}{52}$\omega z = 1+\alpha s$ \end{rotate} }
\psfrag{d2}{\begin{rotate}{52}$\omega z = -1+\alpha s$ \end{rotate} }
\psfrag{d3}{\begin{rotate}{-52}$\omega z = 1-\alpha s$ \end{rotate} }
\psfrag{d4}{\begin{rotate}{-52}$\omega z = -1-\alpha s$ \end{rotate} }
\psfrag{a}{$\frac{1}{\alpha}$}
\psfrag{o}{$\frac{1}{\omega}$}
\psfrag{r}{$s$}\psfrag{z}{$z$}

\subsubsection{The velocity field near the critical latitudes}\label{vfcl}

In order to present in a simple way the singularity of the velocity
field near the critical latitude, we specialize our reasoning to the
case of the tangent characteristic 
\linebreak\mbox{$\omega z = 1-\alpha s$} which touches
the sphere at $s=\alpha$ and $z=\omega$. The velocity component parallel
to this characteristic is such that $V_\parallel=\omega v_s-\alpha v_z$,
or 

\[ iV_\parallel=\frac{\omega^2}{\alpha^2}\ds{P} +
\frac{\alpha}{\omega}\dz{P} \]

\noi Using \eq{trsf2} and \eq{transf_coo2}, we find that

\beqa i\Delta\, V_\parallel &=& (\omega^2\mu\sqrt{1-\eta^2} +
\alpha^2\eta\sqrt{1-\mu^2})\frac{\sqrt{1-\mu^2}}{\alpha}\dmu{P} \\
&& - (\omega^2\eta\sqrt{1-\mu^2}+\alpha^2\mu\sqrt{1-\eta^2})
\frac{\sqrt{1-\eta^2}}{\alpha}\deta{P} \eeqa

Therefore, it turns out that if the right-hand side of this equation
remains finite on the singular surface, then the velocity component
$V_\parallel$ diverges as $1/\Delta = 1/\sqrt{D}$. In the neighbourhood
of the singular surface, $D$ vanishes linearly with the distance to
this surface; thus we see that the velocity field will diverge as one
over square root of the distance to these surfaces as actually found by
\cite{StRic69} in the case of a thin shell.

Let us show that the RHS of the latter equation is indeed finite. The
characteristic $\omega z = 1-\alpha s$ is such that $\mu=\eta$,
therefore

\[ RHS = \mu\frac{(1-\mu^2)}{\alpha}\lp \dmu{P} -\deta{P}\rp_{\mu=\eta} \]

\noi but $\lp \dmu{P} -\deta{P}\rp_{\mu=\eta} =
P'_\ell(\eta)Q_\ell(\eta) - P_\ell(\eta)Q'_\ell(\eta)$ which is the
wronskian of the Legendre functions; it is nonzero as $P_\ell$ and
$Q_\ell$ are linearly independent.

Finally, using the same kind of arguments one may also prove that the
component of the velocity field in the azimuthal direction is also
singular while the component perpendicular to the singular surface
remains finite.

We therefore see that the velocity field possesses an integrable singularity
but is not square-integrable; thus, if strictly quasi-periodic
trajectories of characteristics exist, they would inevitably touch the
critical latitude and their associated eigenfunction would be singular.
Therefore no eigenvalue can be associated with ergodic trajectories of
characteristics in a spherical shell.

\subsection{The toroidal (regular) solutions}

The foregoing two sections have shown us that inertial `modes' of a
sphericl shell hardly escape to singularities: one question therefore
raises up: do regular modes exist at all? the answer is yes, indeed
some regular solutions exist in the form of purely toroidal velocity
fields associated with eigenfrequencies $\omega=1/(m+1), m\in\bbbn^*$.

We pointed out these solutions in \cite{RV97} but they appeared
independently several times in the literature: \cite{Malkus67} noticed
them while investigating hydromagnetic planetary waves and \cite{PaPr78}
called them 'r-modes' because of their similarity with Rossby waves.

However, the existence of these solutions is somehow puzzling since a
plot of the trajectories of characteristics associated with
these eigenvalues shows that most of them converge towards an attractor
(for instance, when $m=2$); how can we reconcile these two apparently
contradictory facts?

The answer lies in the specific form of the Poincar\'e equation in these
cases. Indeed, these modes are purely toroidal which means that for all
points in the spherical shell we have $\er\cdot\vv=0$. From the
expression of the velocity components as a function of the pressure
fluctuation (see for instance \cite{RN99}),  this constraint can be
transformed into the following equation

\[ \frac{\omega}{\alpha^2}s\ds{P} + \frac{mP}{\alpha^2}
-\frac{z}{\omega}\dz{P} = 0 \qquad \forall (s,z) \quad {\rm in\; the\;
domain} \]

\noi When this equation is combined with the Poincar\'e equation, it
turns out that the pressure must satisfy

\beq \frac{\alpha^2}{\omega^2}\lp\frac{z}{s}\dsdz{P} - \ddz{P}\rp
-\frac{m}{\omega s}\ds{P} - \frac{m^2 P}{s^2} =0 \eeq

The characteristics of this hyperbolic equation obey the differential
equation:

\[ zdzds+s(ds)^2 = 0 \]

\noi They are therefore either straight lines parallel to the rotation
axis $ds=0$ or circles parallel to the boundaries $s^2+z^2=K$. They
cannot form orbits by reflections on the boundaries and therefore they do
not impose any constraint on the solution; regular solutions are
possible. In fact, because of the circular shape of one family of
characteristics, the variables of the problem can be separated and solutions
are regular.

Regular inertial modes in a spherical shell therefore exist, but are these
toroidal modes the only regular modes? we have no mathematical proof of
it but numerical computations of the whole spectrum of eigenvalues
including viscosity
strongly suggest that this is indeed the case. The argument is as
follows: regular modes in a spherical shell meeting stress-free
boundary conditions have damping rates proportional to the viscosity
which will turn out to be very small compared to those of singular
modes which, as we shall see, develop shear layers. In a plot of the
eigenvalues in the complex plane, regular modes will pop out when the
viscosity is sufficiently low as can be seen in the context of gravity
modes in \cite{RN99}. Computations for different $m$'s show that only
one eigenvalue popped out and that is the one of the toroidal mode.

\subsection{A summary of the results on the inviscid problem}

Before jumping into the question of how inertial modes of a spherical
shell behave when a slight amount of viscosity is included, it is
certainly useful to summarize the main results obtained in the
foregoing sections on the inviscid problem.

We have seen in \S\ref{stream} that the nature (regular or singular) of
eigenmodes is, with the exception of toroidal modes, determined by the
dynamics of characteristics. The study of this dynamics
(\S\ref{dyn_char}) revealed the generic property that characteristics
converge to attractors made of a periodic orbit which exist in some
frequency band. These attractors are also characterized by their length
(\ie the number of reflexions) which influence the rate at which
characteristics converge to them; this rate is given by a negative
Lyapunov exponent. When the frequency of the attractor is close to
$\sin(p\pi/2q)$ where $p$ and $q$ belong to a finite set of integers
determined by the size of the inner core, the corresponding attractors
are very long and weakly attractive. These frequencies are the ones for
which the shadow of the inner core follows a periodic orbit (see
figure~\ref{fig_shad}); they will prove to be important in the
determination of the asymptotic spectrum of inertial modes when the
viscosity vanishes.

The focusing of energy by attractors is not the only source of
singularity: we have shown that near the critical latitude of the inner
boundary a milder
singularity will develop in general. This singularity will prove to be
relevant in the viscous case when shear layers associated with
attractors are inhibited.

Finally, we found that some regular solutions still exist. They
are purely toroidal modes and we surmize that they are the only true
eigenmodes of a rotating fluid in a spherical shell. From the
mathematical point of view, the point spectrum of the Poincar\'e operator (\ie
eigenvalues associated with square-integrable functions) is almost
empty.

\section{The solutions with viscosity}

\subsection{General results}\label{thex}

When viscosity is included the equations are elliptic and the problem
is well-posed; hence, the solutions are smooth ${\cal
C}^\infty$-functions which can be computed numerically. We shall not
describe the method used and refer the reader to
\cite{RV97}. We just recall that we solve the eigenvalue problem
($\lambda$ is the complex eigenvalue)

\greq
\lambda\vu+\ez\times\vu = -\na P + E\lap\vu\\
\na\cdot\vu = 0
\egreqn{eqmot}
with stress-free boundary conditions to eliminate Ekman boundary
layers; $E=\nu/2\Omega R^2$ is the Ekman number, $\nu$ being the
kinematic viscosity.

The main result, coming from numerical solutions of this problem, is that
the amplitude of the modes is concentrated along paths of characteristics
drawn by attractors. However, as found by \cite{DRV99} and \cite{RV97},
the attractor appears in the viscous solutions only when
the Ekman number is low enough. This critical Ekman number, below which
the mode seems to reach an asymptotic shape, depends on the length of
the attractor; short (and simple) attractors appear at higher
viscosities than long (and complex) attractors which may never appear
within the range of physically relevant Ekman numbers ($E\supapp
10^{-18}$).

In order to investigate the properties of viscous solutions associated with
attractors, we shall focus on a few simple ones which appear at
reasonable Ekman numbers (\ie larger than $10^{-9}$). Some are the ones
displayed in figure~\ref{orbs} plus two others located in the
0.6--0.625 frequency band, one of which was considered by \cite{Israe72}
using a thin shell.

\begin{figure}
\centerline{\includegraphics[width=7.5cm,angle=0]{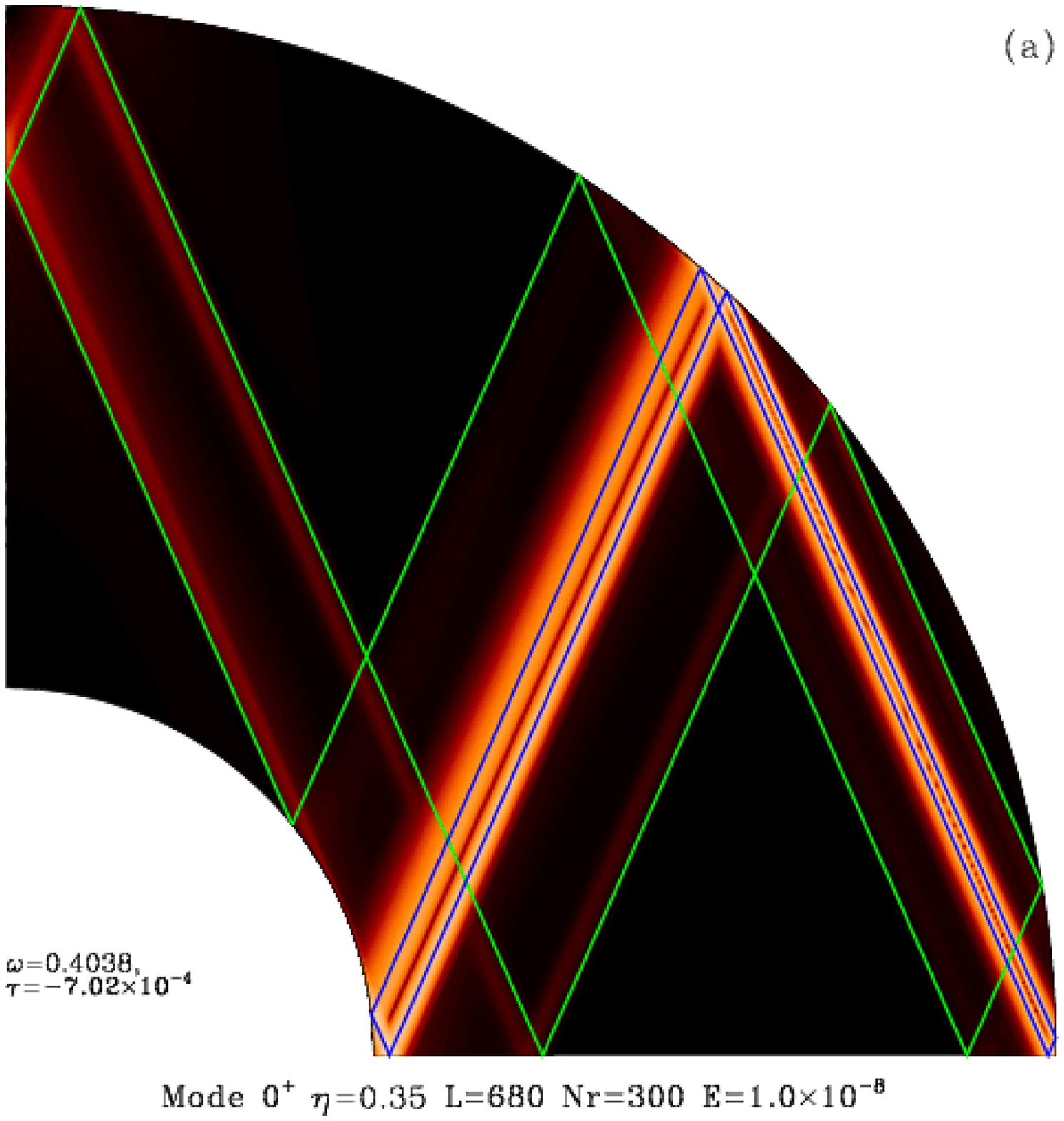}
\includegraphics[width=7.5cm,angle=0]{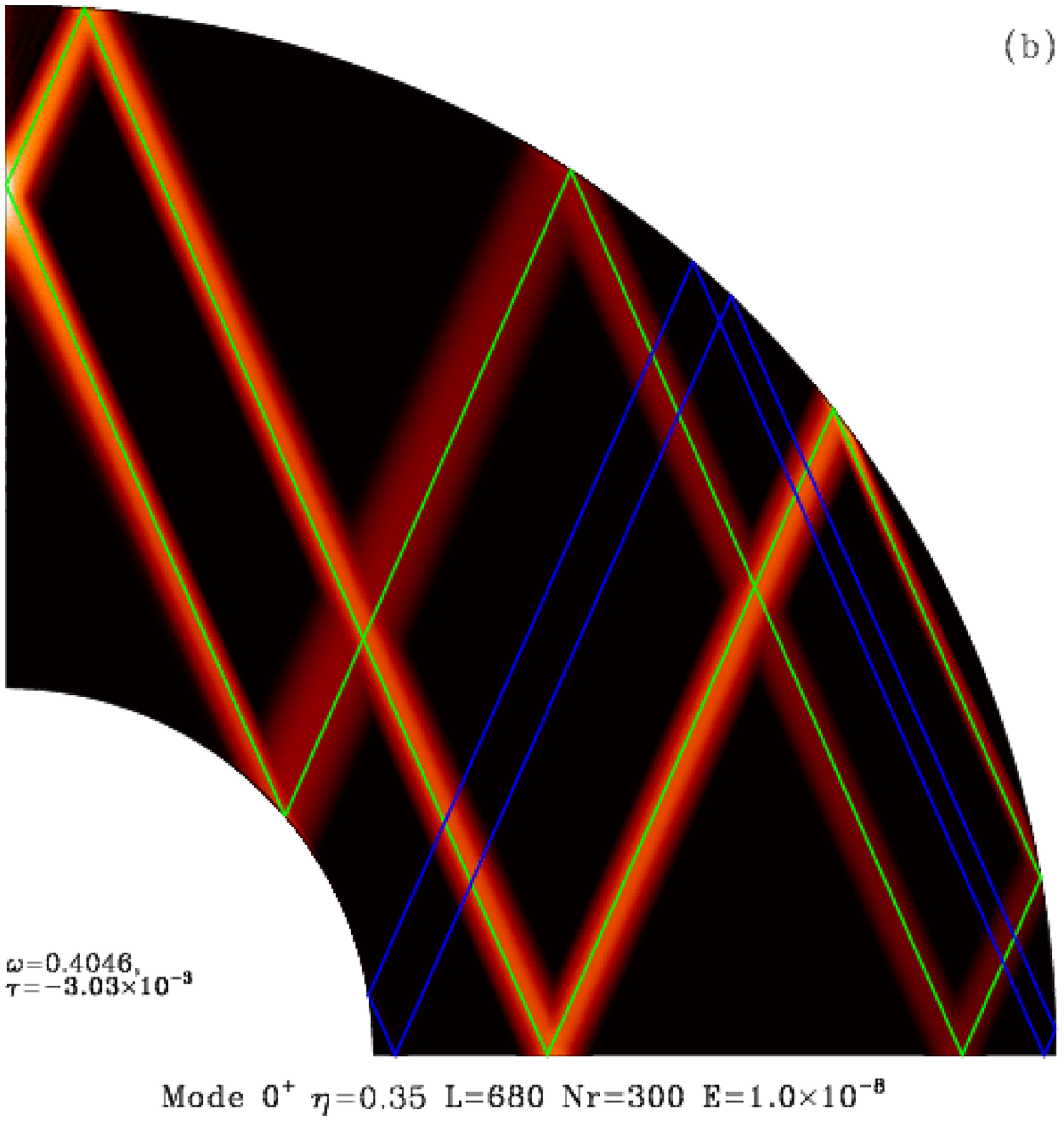}}
\centerline{\includegraphics[width=7.5cm,angle=0]{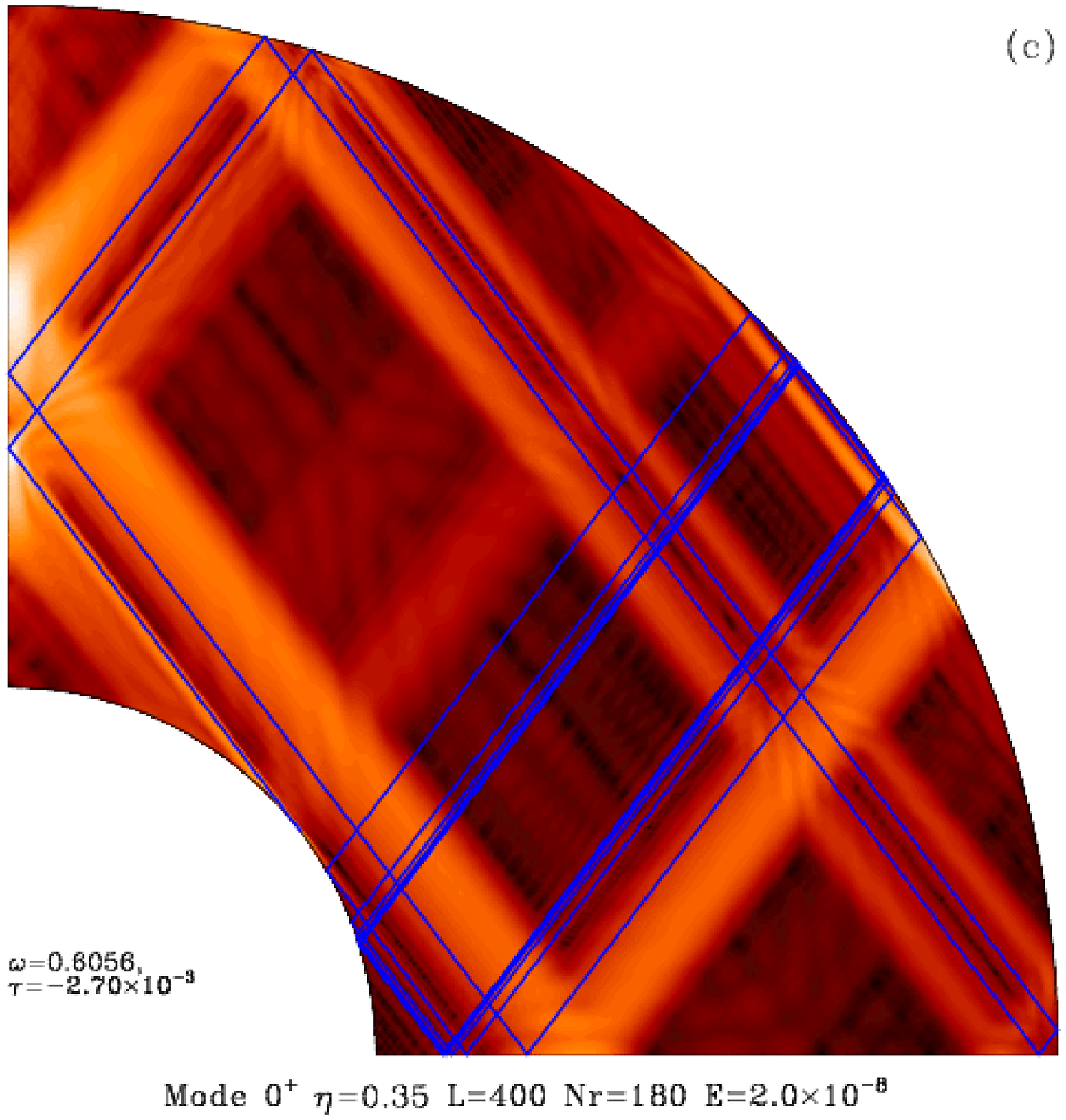}
\includegraphics[width=7.5cm,angle=0]{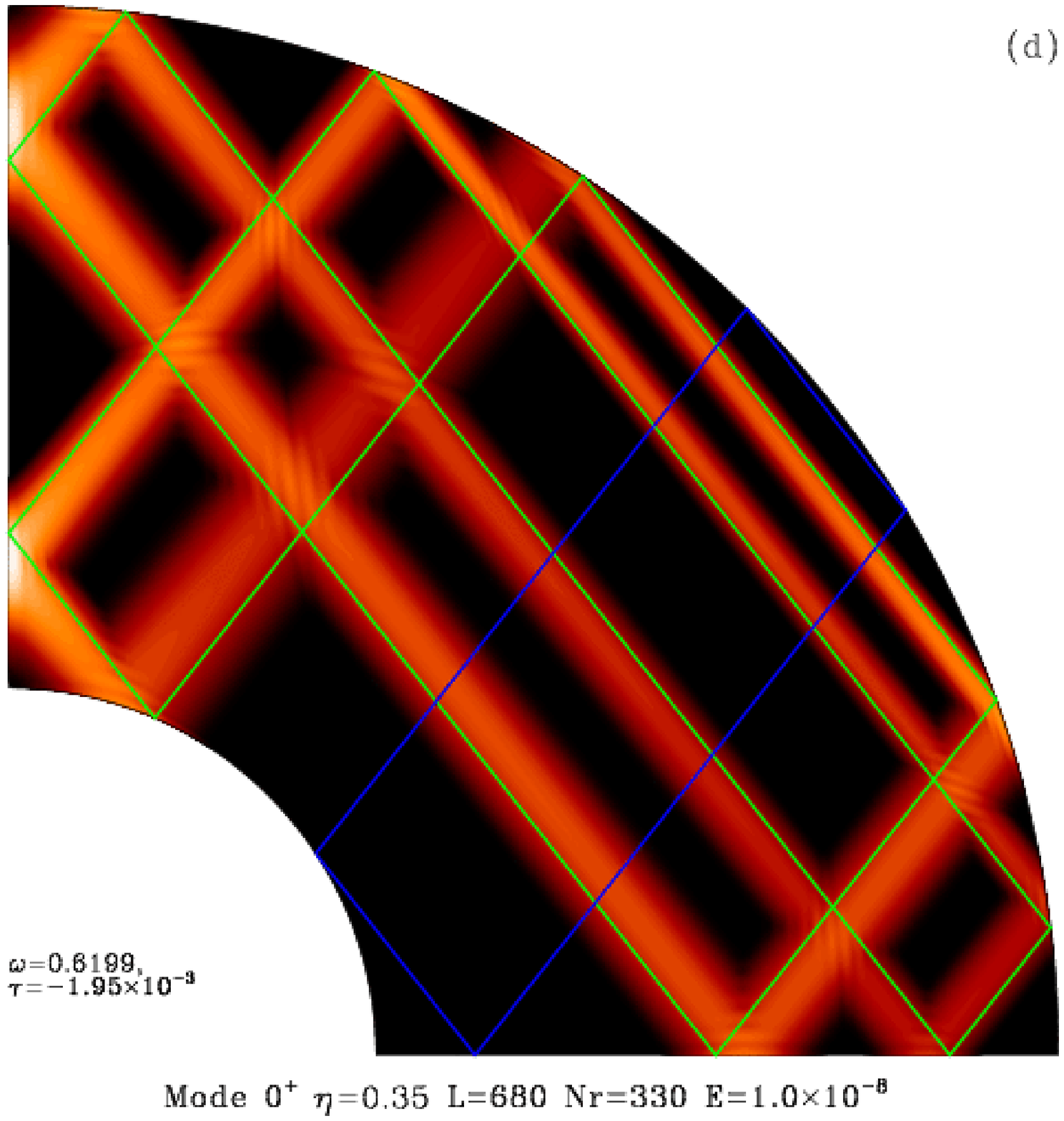}}
\caption[]{Distribution of kinetic energy in the meridional section of
the shell for four different axisymmetric modes. These solutions have been
computed
numerically using the same code as in \cite{RV97}. On each panel, $\tau$
is the damping rate, stress-free boundary conditions are used, $L$ is
the number of spherical harmonics and Nr is the number of grid points in
the radial direction. (a) shows the
mode associated with the equatorial attractor (in blue) of figure~\ref{orbs}a
while the polar attractor (in green) is only weakly visible. (b) Using
a more damped mode
with a slightly different frequency, we obtain a case where the polar
attractor is fueled with energy. In (c) the attractor considered by
\cite{Israe72} should be fueled, but is in fact hampered by the
boundary and the critical latitude singularity dominates the flow.
(d) A very neat mode which concentrates along its attractor; in blue the
corresponding equatorial attractor.}
\label{num_sol}
\end{figure}

 A plot of the eigenmodes associated with these four attractors is
shown in figure~\ref{num_sol}. This figure displays the kinetic energy
of the modes in a meridional section of the shell. As expected the
kinetic energy focuses around the attractors which we overplot on each
diagram; however, this is not systematic as shown by
figure~\ref{num_sol}c: there, the kinetic energy concentrates along a
characteristic path starting at the critical latitude rather than along
the (only) existing attractor. We understand this situation as the
consequence of the location of the attractor: one of its segments is
indeed almost tangential to the outer sphere which therefore inhibits
the development of the shear layer. By computing the distance between
the boundary and the attractor, we estimate that a $E^{1/4}$-shear
layer is inhibited by the boundary if the Ekman number is larger than
$10^{-11}$.  Such low Ekman numbers are out of reach numerically at the
moment.  This example illustrates the point mentioned in
\S\ref{sect_attr}:  very long attractors will appear at extremely low
Ekman numbers. If we consider that lowest Ekman numbers are those of
stars ($\sim 10^{-18}$) or the Earth's core ($\sim 10^{-16}$) and if we use the same
scaling as above for shear layers, then we can conclude (from
figure~\ref{period_distance}) that attractors longer than $\sim 100$ will
never appear in physical systems.

We therefore see that, although the singularity associated with an
attractor is stronger than that of the critical latitude, this latter
singularity may show up if, for some reason, the build up of shear
layers around the attractor is inhibited. We surmise that `long'
attractors, will dominate relative to the critical
latitude singularity only at very low Ekman numbers. `Short' attractors
may therefore appear more easily as in figure~\ref{num_sol}a while still
showing, weakly, the critical latitude singularity.

Another surprising feature of the rays (\ie shear layers)
lying along an attractor is that
the maximum energy density is not always centered on the attractor
(figure~\ref{num_sol}b or \ref{num_sol}d). We discuss this point below.

To make some progress in the understanding of this complex behaviour we
shall investigate in more detail
the structure of shear layers lying near the attractors.

\subsection{Structure of shear layers}

\subsubsection{Some numerical results}

\begin{figure}
\centerline{\includegraphics[width=8cm,angle=0]{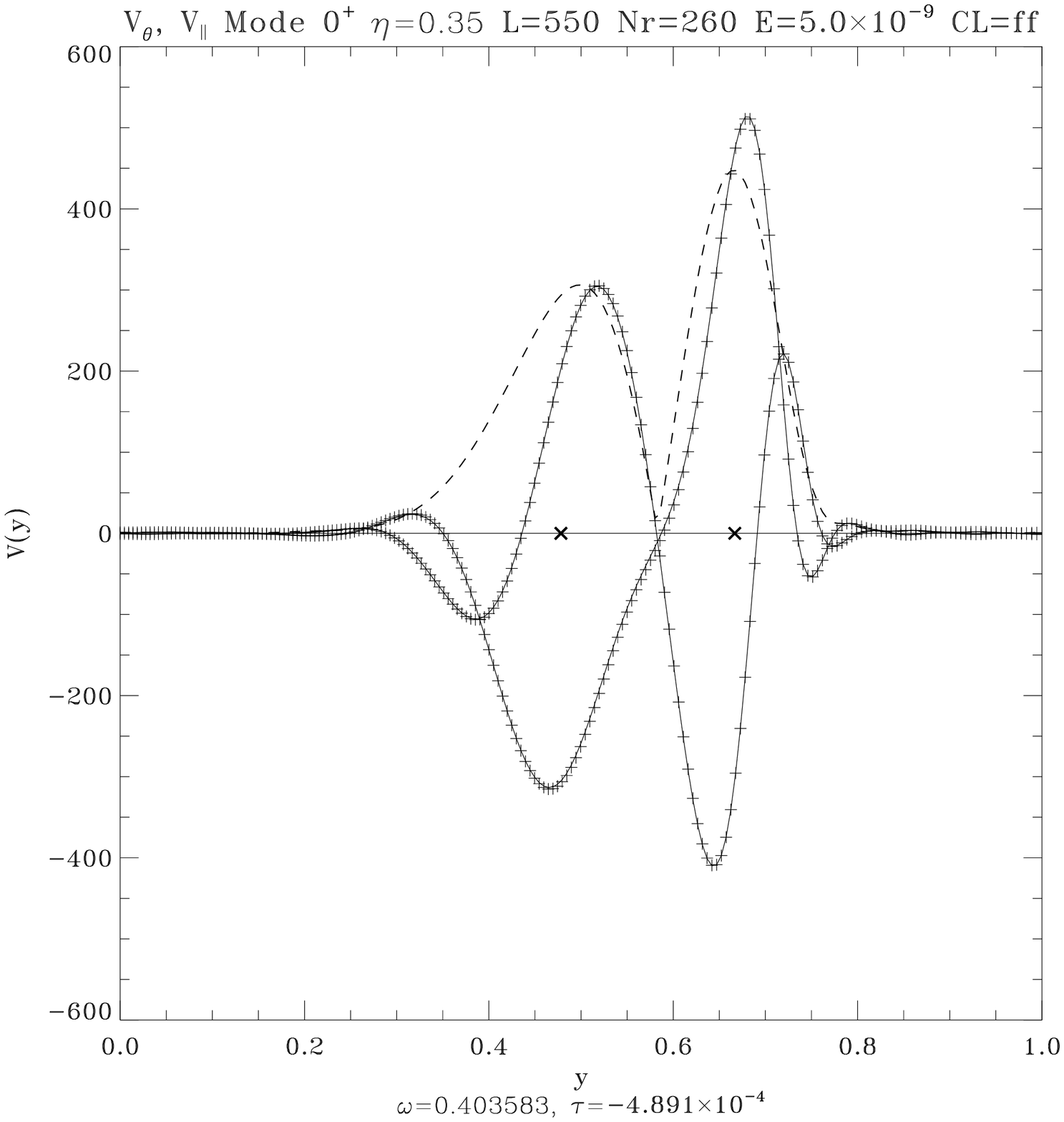}
\includegraphics[width=8cm,angle=0]{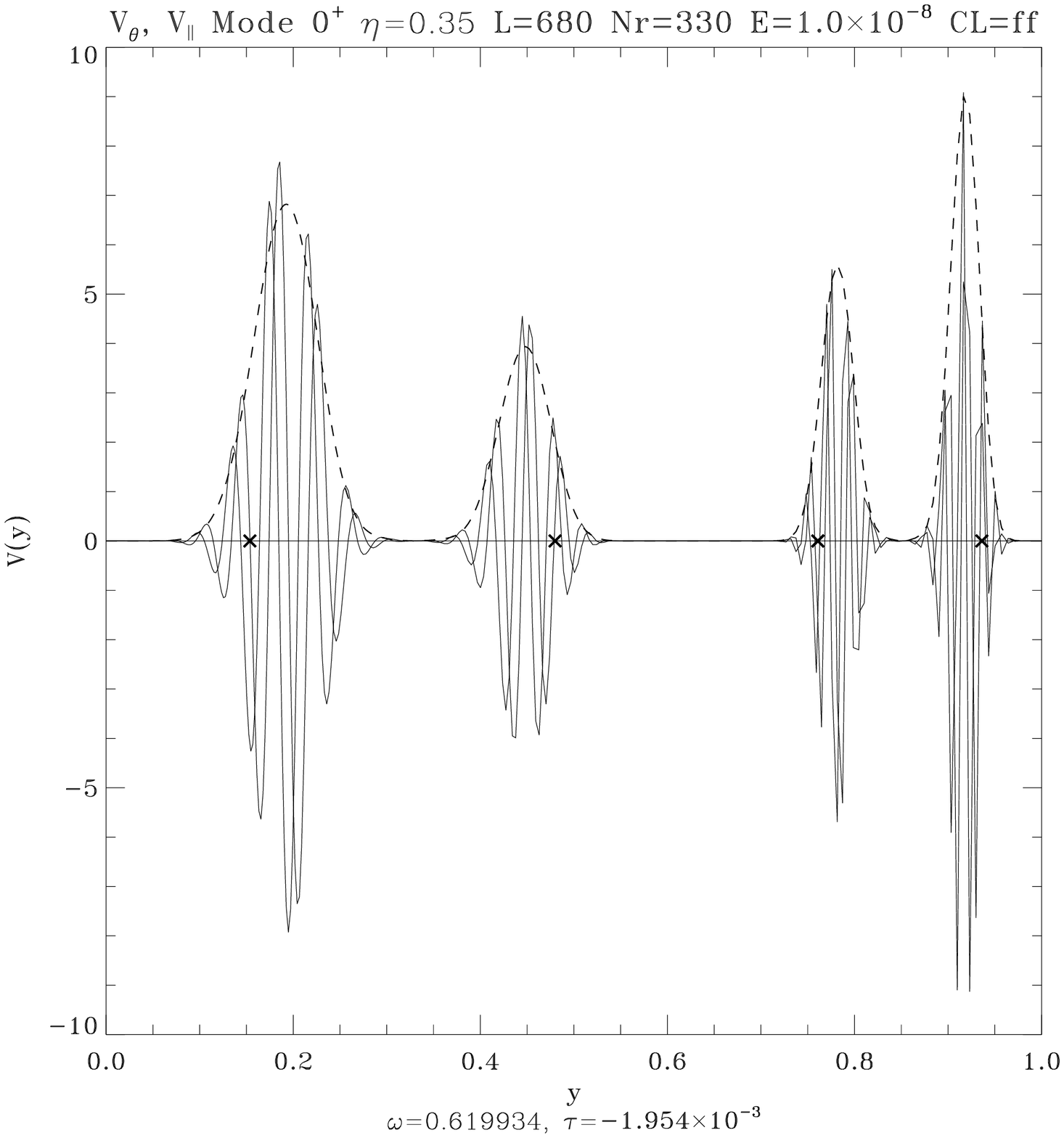}}
\caption[]{Left: the real and imaginary parts of
$V_\varphi$ in the cross-section of the attractor displayed in
figure~\ref{num_sol}a. The segments with positive slope have been shown in
cross section.  The $+$ sign overplotted on the curves represents the
variations of $iV_\parallel$; the perfect matching with the curves of
$V_\varphi$ shows that the phase quadrature between these components is
well verified as expected from \eq{phase_quad1} or \eq{phase_quad2}. The
dashed line shows the amplitude of the $|\vv|$ profile and the two crosses
on the $y$-axis indicate the position of the attractor. Right: same as on left
but for a mode with more complex rays: it is a cut through the rays with
negative slope of the mode of figure~\ref{num_sol}d; the cut starts
near the critical latitude and is perpendicular to the rays.}
\label{shear_layer}
\end{figure}

As a preliminary step, we first compute the variations of
the components of the velocity field along a line crossing a ray
perpendicularly. Results are displayed in figure~\ref{shear_layer}.

These profiles show that these internal shear layers have a rather
complex structure which looks like a plane inertial wave trapped in a
`potential well'. Each mode seems to be characterized by the number of
nodes in the cross-section of its rays, just like a solution of a
Sturm-Liouville problem. The analogy cannot, however, be pushed too far
since the actual oscillations do not disappear outside the rays but
continue with a very low amplitude (the well is leaking!).
This is a consequence of the fact that the `well' is not a local well
but the result of a mapping made by the convergence of
characteristics towards the attractor. We also surmise that since the
convergence rate is not the same on each side of the
attractor\footnote{We mean here at some finite distance from the
attractor; right on the attractor the convergence rate is given by the
Lyapunov exponent.}, the `potential well' is certainly not symmetric
with respect to the attractor; we thus explain our finding that the maxima
of kinetic energy density are not centered right on the attractor as
shown by figures~\ref{num_sol}b,d or \ref{shear_layer}.

In the above view, the shear layers
result from a balance of the focusing action of the mapping and the
`defocusing' action of viscosity; because the former action is global
and the latter is local, the boundary layer analysis is difficult, if not
impossible. The following analysis gives some general properties of
these shear layers, properties which are actually observed numerically,
but is not able to reproduce their detailed structure.

\subsubsection{Boundary layer analysis}

In order to describe the shear layers featuring the inertial modes of a
spherical shell, it is convenient to project the equations in the
characteristics' directions.

From \eq{charac}, we derive the expressions of unit vectors parallel
($\parallel$) or perpendicular ($\perp$) to characteristics of positive
($+$) or negative ($-$) slope:

\[ \epar_\pm = \omega\es\pm\alpha\ez,  \qquad
\eper_\pm = \omega\ez\mp\alpha\es \]

Restricting ourselves to the case of a characteristic with positive slope, the
components of the velocity in a meridional plane are:

\[ V_\parallel = \omega V_s + \alpha V_z, \qquad
V_\perp = - \alpha V_s + \omega V_z \]

We may now transform the equations of motions, written in cylindrical
coordinates,

\greq
\disp{\lambda V_s - V_\varphi = -\ds{P} + E\Delta'V_s}\\
\\
\lambda V_\varphi + V_s = E\Delta'V_\varphi\\
\\
\disp{\lambda V_z  = -\dz{P} + E\lap V_z}
\egreq

\noi into

\greq
\disp{\lambda V_\parallel - \omega V_\varphi = -\dx{P} + E\lp\lap -
\frac{\omega^2}{s^2}\rp V_\parallel+ \frac{\alpha\omega E}{s^2}V_\perp}\\
\\
\lambda V_\varphi + \omega V_\parallel -\alpha V_\perp =
E\lap V_\varphi\\
\\
\disp{\lambda V_\perp + \alpha V_\varphi = -\dy{P} + E\lp\lap+
\frac{\alpha^2}{s^2}\rp V_\perp -\frac{\alpha\omega E}{s^2}V_\parallel}
\egreqn{dyneq2}

\noi where $\Delta' = \lap -1/s^2$, and $x$ and $y$ are local coordinates
respectively parallel and perpendicular to the characteristic. For the
sake of simplicity, we may consider one such characteristic so that

\beq x=\alpha z +\omega s, \qquad y=\omega z - \alpha s, \qquad 
s= \omega x -\alpha y\eeq

\noi Thus

\[ \ds{} = \omega\dx{} - \alpha\dy{}, \qquad \dz{} =
\alpha\dx{}+\omega\dy{} \]

Mass conservation requires that

\beq \dx{V_\parallel} + \dy{V_\perp} + \frac{\omega V_\parallel - \alpha
V_\perp}{s(x,y)} = 0 \eeqn{mass}

As it was shown in \cite{RV97}, the inviscid balance along rays shows a
dependence of the velocity and pressure fields with $1/\sqrt{s}$; we
shall remove such a dependence from our equations by setting
$\vV=\vu/\sqrt{s}$ and $P=p/\sqrt{s}$. Hence, \eq{dyneq2} and \eq{mass}
yield

\greq
\disp{\lambda u_\parallel - \omega u_\varphi = -\dx{p} +\frac{\omega
p}{2s} + E\lp\lap
-\frac{\omega^2}{s^2}\rp u_\parallel+ \frac{\alpha\omega
E}{s^2}u_\perp}\\
\\\lambda u_\varphi + \omega u_\parallel -\alpha u_\perp =
E\lap u_\varphi\\\\
\disp{\lambda u_\perp + \alpha u_\varphi = -\dy{p} -\frac{\alpha p}{2s} +
E\lp\lap+\frac{\alpha^2}{s^2}\rp u_\perp -\frac{\alpha\omega
E}{s^2}u_\parallel}\\ \\
\disp{\dx{u_\parallel} + \dy{u_\perp} + \frac{\omega u_\parallel -
\alpha u_\perp}{2s} = 0}
\egreqn{dyneq3}

\noi where now $\lap=\dds{}+\frac{1}{4s^2} + \ddz{}=
\ddx{}+\ddy{}+\frac{1}{4(\omega x-\alpha y)^2}$.

\subsubsection{The inner E$^{1/3}$-layer}\label{1tiers}

Searching for a boundary layer solution scaling with $E^{1/3}$, we make
the expansion

\greq
u_\parallel = u_0^\parallel + E^{1/3}u_1^\parallel + \cdots \\
u_\varphi = u_0^\varphi + E^{1/3}u_1^\varphi + \cdots \\
u_\perp =  E^{1/3}u_1^\perp + \cdots \\
p =  E^{1/3}p_1 + \cdots 
\egreq

\begin{figure}
\centerline{\includegraphics[width=8cm,angle=0]{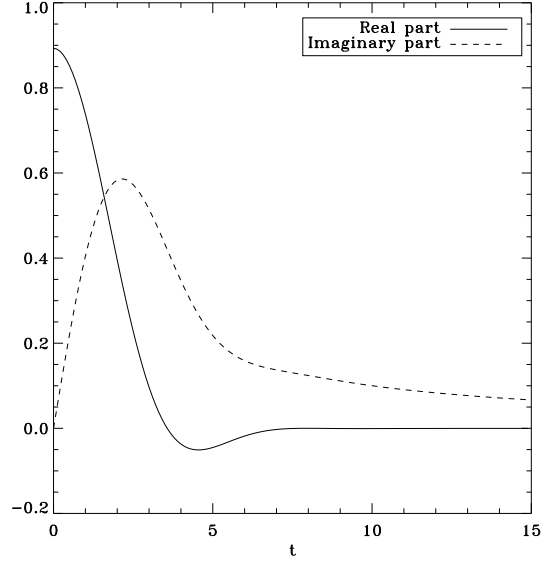}}
\caption[]{Shape of the Moore-Saffman function with $m=-1/3$.}
\label{MS_func}
\end{figure}

\noi We shall use the scaled variable $Y=y/E^{1/3}$. We also recall 
that $\lambda=i\omega+\tau$, $(\omega,\tau)\in\bbbr^2$ and that
$|\tau|\ll |\omega|$. From
the second and third equations of \eq{dyneq3} at zeroth order we get

\beq u_0^\varphi = i u_0^\parallel \qquad {\rm and}\qquad \alpha u_0^\varphi
= -\dY{p_1}\eeqn{phase_quad1}

\noi while from the combination of first order terms we get

\beq \dddY{u_0^\varphi} = -i\alpha\dx{u_0^\varphi} \eeq

\noi Making a last change of variables $q = x/\alpha $ and dropping the
zero-index, we finally obtain

\beq \dddY{u_\varphi} = -i\dq{u_\varphi} \eeqn{MS}

\noi which was first derived by \cite{MS69} for steady (vertical) shear
layers.

Moore and Saffman have shown that the solutions of \eq{MS} which can
describe a detached shear-layer are self-similar solutions of the form:

\beq u_\varphi = q^m H_m\lp Y/q^{1/3}\rp \eeqn{solMS}

\noi where the function $H_m$ is defined by

\[ H_m(t) = \int_0^\infty e^{-ipt}e^{-p^3}p^{-3m-1} dp \]

\noi Since \eq{solMS} describes detached shear layers, the function
$H_m$ needs to vanish when $t \rightarrow \pm\infty$ which is possible
only if $m<0$ \cite[]{MS69}. The shape of this function is given in
figure~\ref{MS_func}.

\psfrag{a}{$A$}\psfrag{ap}{$A'$}\psfrag{b}{$B$}\psfrag{c}{$C$}
\psfrag{bul}{$\bullet$}
\psfrag{bulb}{\hspace*{-0.5mm}\raisebox{-.3mm}{$\bullet$}}
\psfrag{bulc}{\hspace*{0.5mm}\raisebox{0.5mm}{$\bullet$}}
\psfrag{ell}{$\ell$}
\begin{figure}
\centerline{\includegraphics[width=8cm,angle=-90]{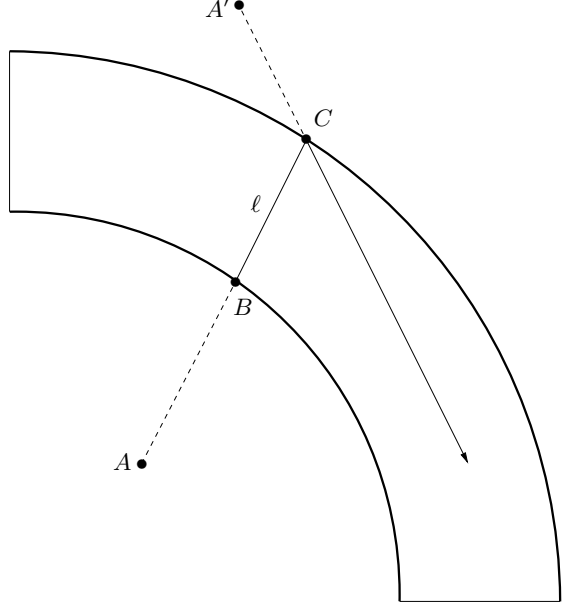}}
\caption[]{Motion of the virtual source after a reflection: the source
$A$ moves to $A'$ after the reflection of the ray.}
\label{source_virt}
\end{figure}

To complete the description of the $\frac{1}{3}$ layer we need to
determine the index $m$ of the Moore and Saffman function. For this
purpose, we first note from \eq{solMS} that the width of the layer
is singular at the origin of the $x$-axis. This origin can therefore be
considered as a virtual source of the ray which obviously lies outside
the fluid's container. Let us therefore consider a segment of a mode around
an attractor. Let us orient the $x$-axis in the direction of contraction
of the map and call $x_0$ the abscisse of the first point of this
segment (point $B$ in figure~\ref{source_virt}). At $B$ the width of the
layer is proportional to $x_0^{1/3}$ while the amplitude of $u_\varphi$
is proportional to $x_0^mH_m(0)$. At $C$ and before reflection, the width is
$(x_0+\ell)^{1/3}$ and the amplitude is $(x_0+\ell)^mH_m(0)$; after
reflection, the mapping changes  the scale by a factor $K$; therefore the
width is now $(x_0+\ell)^{1/3}K$ and the amplitude
$(x_0+\ell)^mH_m(0)/K$. The width is just as though the virtual source
$A'$ were at a distance $(x_0+\ell)K^3$ of $C$ while the amplitude would
imply a distance $(x_0+\ell)K^{-1/m}$; since the segment starting at $C$
must also be a solution of the form \eq{solMS}, the new virtual source
must be at the same position for both the amplitude and width; therefore,
we need to have

\beq m=-\frac{1}{3} \eeq

This index was also found by \cite{stewar72a} on the argument that it
is the only one for which the flux 

\[ \int_{-\infty}^{+\infty} V^\parallel dY \]

\noi is conserved along the ray, which means that the $\frac{1}{3}$
layer does no pumping.

From the property that $H_m(t)\sim t^{3m}$ as $t\rightarrow\infty$, we
see that the solution \eq{solMS} is independent of $x$ far from the
layer and decreases as $1/y$.

\subsubsection{The outer E$^{1/4}$-layer}

Since some modes show a clear scaling of their ray with the
$\frac{1}{4}$ exponent \cite[see][]{RV97}, we also briefly discuss this
case.

As the $\frac{1}{4}$-layer is also much larger than the Ekman layer, the
$\varphi$- and $\parallel$-components are in quadrature, \ie,

\beq V_\varphi = iV_\parallel \eeqn{phase_quad2}

\noi These components also verify, from mass conservation and inviscid
balance,\linebreak $V_\varphi=F(y/E^{1/4})/\sqrt{s}$.

We cannot say much more about the $\frac{1}{4}$-layer, except that the
expansions to higher orders including viscous terms lead to a
differential equation for $F$ which is not closed, some additional
functions missing their differential equation. If these extra and
undetermined functions are set to zero, $F$ obeys, as expected, a
fourth order differential equation whose solutions do not agree (in
general) with the numerical results. We think that this is due to
the action of the mapping which is obviously missing in the local
analysis; unfortunately, we have not yet found a way to include it.

\subsubsection{Wave packet kinematics}

In order to give a more physical understanding of
the behaviour of shear layers as viscosity is reduced, we propose
to consider a wave packet traveling around an attractor.

Let us suppose that the Ekman number is very small but finite. When
traveling along an attractor a wave packet is damped by viscosity but
its reflections on the boundaries enhance it if the direction of
propagation is such that the map is contracting ($\Lambda<0$); this
equilibrium may be written as:

\beq e^{-\nu\sum_n k^2_nt_n} = e^{N\Lambda} \eeqn{equil_amp}

\noi where $N$ is the length of the attractor, $\Lambda$ its Lyapunov
exponent\footnote{recall that according to the definition of the
Lyanunov exponent, $e^\Lambda$ is the mean dilation rate (of an
interval $\delta\phi$) per bounce.}, $t_n$ the time elapsed on the 
n$^{th}$-segment and $k_n$ the
wavenumber of the n$^{th}$-segment. Noting that on each segment the group
velocity is almost constant and reads:

\[ v_g = 2\Omega\frac{k_s}{k^2} \]

\noi we may transform \eq{equil_amp} into

\beq \Lambda = -\frac{E}{N\sqrt{1-\omega^2}}\sum_n k^3_n \ell_n
\eeqn{rel1}

\noi where we have introduced the length $\ell_n$ of each segment of the
attractor; we also used the fact that $k_s=k\sqrt{1-\omega^2}$ so that
$t_n=\ell_nk_n/\sqrt{1-\omega^2}$ and all quantities are now
dimensionless. Since
$k_n=C_nk_{n-1}$, $C_n>1$ being the contraction coefficient of the n$^{\rm
th}$ reflection, we may rewrite \eq{rel1} 

\beq \Lambda = -\frac{Ek_1^3}{N\sqrt{1-\omega^2}}
\sum_{n=1}^N \ell_n\prod_{i=2}^n C_i^3 = - Ek_1^3 F(\omega) \eeqn{rel1p}

\noi where

\[ F(\omega) = \frac{1}{N\sqrt{1-\omega^2}}\sum_{n=1}^N
\ell_n\prod_{i=2}^n C_i^3 \]

\noi is a purely geometrical quantity describing the path of characteristics
associated with the attractor at the frequency $\omega$. The expression
\eq{rel1p} expresses through \eq{equil_amp} the strict periodicity of
the amplitude of the velocity field along an attractor when viscosity is
small but finite.

A similar relation may be derived if we now express that the scale of a
wave packet must be the same after one cycle along the attractor. In a
purely diffusive (viscous) process, the scale of a structure, initially
being `$a$' grows like $\sqrt{a^2 + \nu t}$ with time; therefore the
relation between the layer's width after one propagation and one
reflection is

\beq a_n = D_n\sqrt{a_{n-1}^2 + \nu t_{n-1}} \eeqn{recur}

\noi Here $D_n$ is the dilation coefficient of the  
n$^{\rm th}$-reflection ($D_n=1/C_n<1$). Along one cycle with N
reflections, we have 

\[ \Lambda  = \frac{1}{N} \ln \prod_{n=1}^N D_n \]

\noi Using the fact that $a_{N+1}=a_1$,  \eq{recur} leads to:

\beq \Lambda =-\frac{1}{2N}\sum_{n=1}^N\ln\lp 1 +\frac{\nu
t_n}{a_n^2}\rp \eeq

Now, if we let the width of rays $a_n$ scale with $E^\sigma$, we find that
$\nu t_n/a_n^2 \sim E^{1-3\sigma}$; imposing $0<\sigma<1/3$, we finally
obtain

\[ \Lambda = -\frac{E}{2N\sqrt{1-\omega^2}} \sum_n \frac{\ell_n
k_n}{a_n^2} \]

\noi Noting that $a_n \sim \lambda_n = 2\pi/k_n$, we recover \eq{rel1} except
for a constant factor.

The two derivations of the Lyapunov exponent through this schematic
model show that the width of shear layers lying along a periodic
attractor and scaling with $E^\sigma$, should be such that 

\[ \sigma < \frac{1}{3} \]

\noi We therefore see that the $\frac{1}{3}$-exponent is a limit case.
In fact this inequality shows that `naked' $\frac{1}{3}$-layers cannot
exist and should be embeded in thicker layers; this
seems to be the case indeed, at least for all the modes which we
investigated in detail: they usually show $\sigma\simeq 1/4$.

\begin{figure}
\centerline{\includegraphics[width=0.5\linewidth]{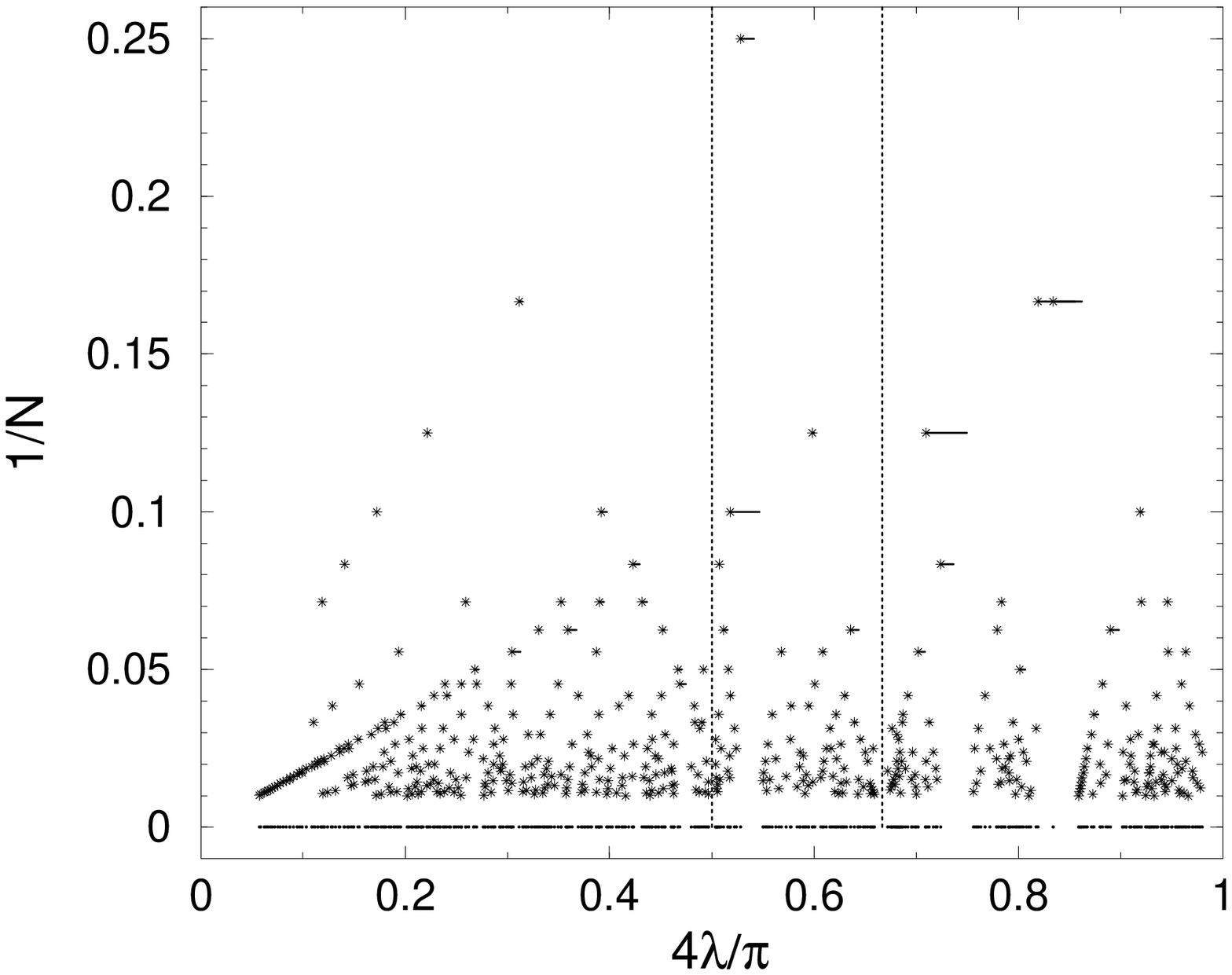}
\includegraphics[width=0.5\linewidth]{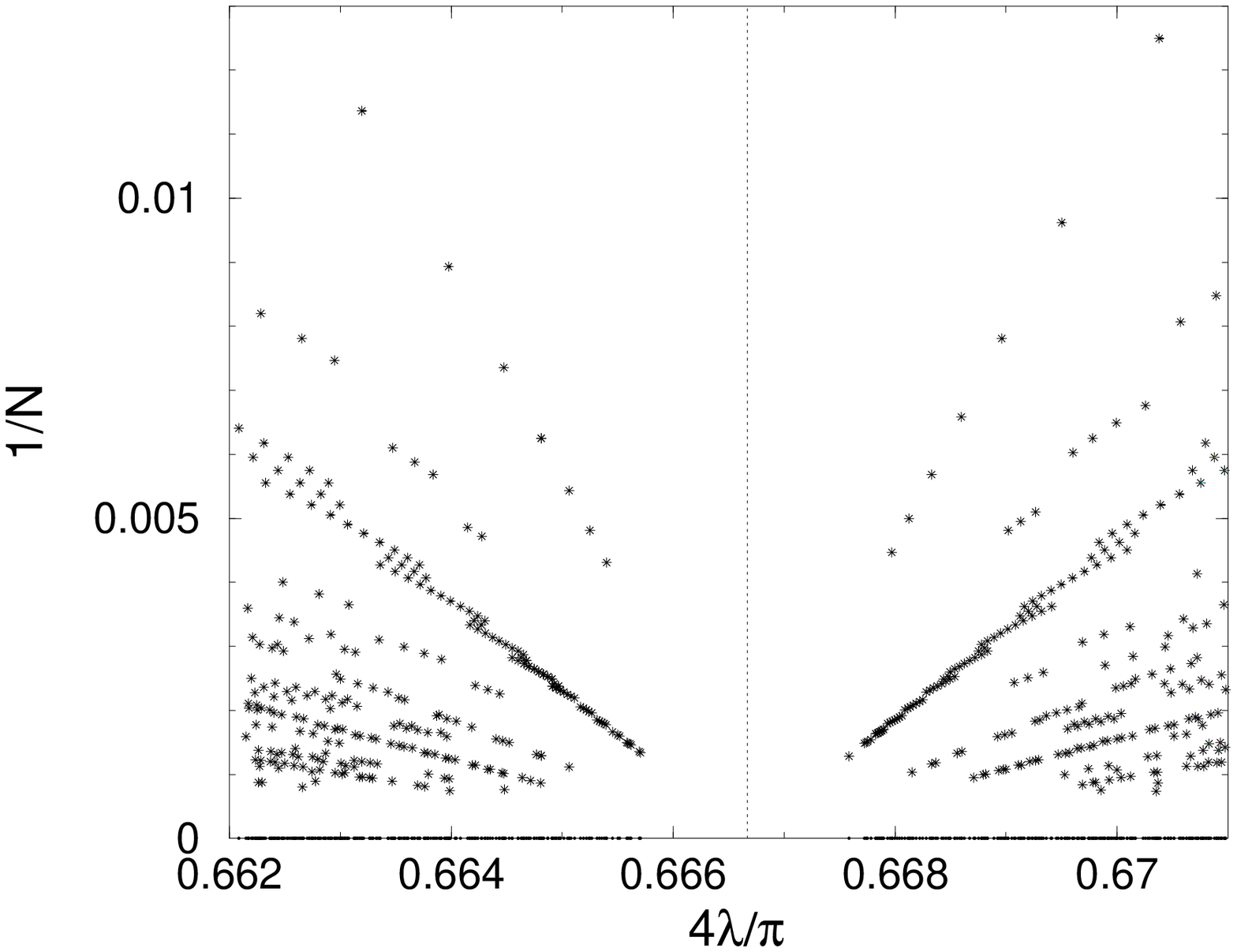}}
\caption[]{Left: The inverse of the length $N$ of all the attractors with a
length less than 100 for a spherical shell with $\eta=0.35$ are
represented with a (*) denoting the frequency (or critical latitude)
where $\Lambda=0$ and with a line segment showing the interval of
existence. By showing the projection of the * on the x-axis, we try to
give an idea of what would be the asymptotic spectrum neglecting very
long attractors. The two vertical lines show the position of $\pi/8$
and $\pi/6$ which correspond to the periodic orbits of the shadow for
$\eta=0.35$. Right: A blow-up of the region around $\pi/6$; note the
lengthening of the attractors as this value is approached.}
\label{asymp_spectrum}
\end{figure}

\subsection{The asymptotic spectrum}

The foregoing calculations show one important result: As viscosity
tends to zero and since $k_1 \propto E^{-\sigma}$, from~\eq{rel1p}
we may conclude that the Lyapunov exponent of attractors
must vanish as viscosity vanishes following the law 
$\Lambda\propto E^{1-3\sigma}$. It therefore turns out that 
eigenfrequencies will
converge towards the roots $\omega_i$ of the equation
\mbox{$\Lambda(\omega)=0$} which therefore describe the asymptotic spectrum of
inertial modes in a spherical shell. From the fact that only a finite
number of attractors exist at a given frequency, we deduce that the
spectrum cannot be dense in [0,1] contrary to the
case of the full sphere. However, this spectrum has some accumulation
points $\omega_a$ which are due to the existence of neutral
\mbox{($\Lambda=0$)} periodic orbits with frequencies $\sin p\pi/2q$ (see
\S\ref{FSC}). Indeed, in the neighbourhood of such points we may find
attractors which are longer and longer, the closer they are to
$\omega_a$. However, the number of accumulation points is finite
and given by the number of pairs $(p,q)$ possible for periodic
orbits of the shadow and when $\eta \geq 1/\sqrt{2}$ only three such
points ($\omega=0,1/\sqrt{2}, 1$) exist. In figure~\ref{asymp_spectrum}
we clearly see the accumulation points corresponding to $\sin(\pi/4)$
and $\sin(\pi/6)$\footnote{The case $\sin(\pi/8)$ is not as clear for
it needs a much higher frequency resolution since the shadow almost
fills the whole volume as $\pi/8 \sim \arcsin(0.35)$.}.
When $\eta \rightarrow 0$, the number of
accumulation points gets larger and larger, as more and more rationals
are added to the set of accumulation points, a situation in accordance
with the fact that at $\eta=0$ the spectrum is dense in $[0,1]$.

Let us also underline the fact that when $E=0$, eigenvalues disappear
since solutions of the equations are no longer square-integrable; this
is also true for frequencies of attractors such that $\Lambda=0$, since
attractors still focus the energy (but algebraically, not exponentially).

\begin{figure}
\hfil
\begin{minipage}[t]{.46\linewidth}
 \centerline{\includegraphics[width=1.1\linewidth]{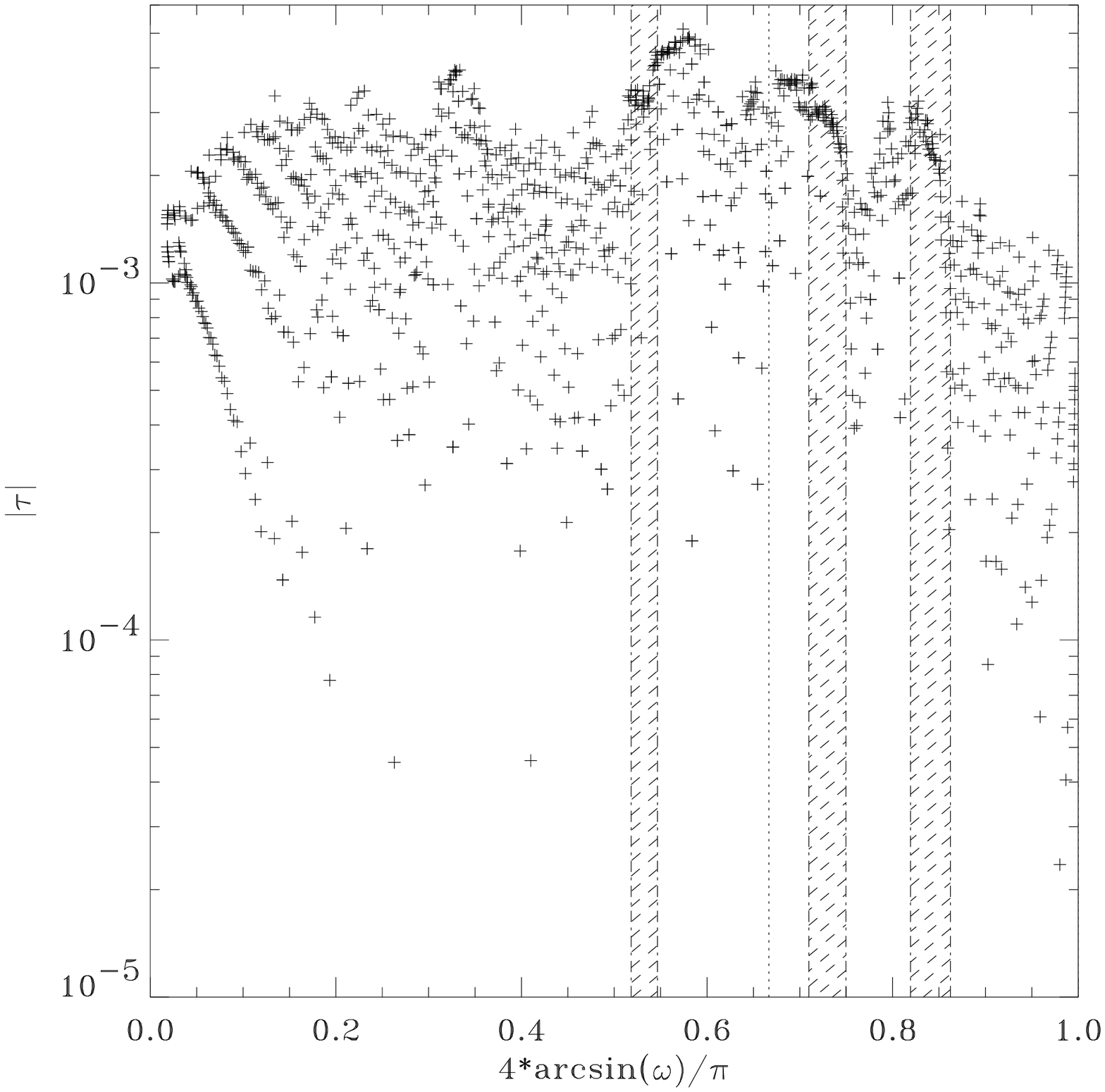}}
\caption[]{Distribution of the least-damped eigenvalues in the complex
plane when $E=10^{-8}$ with resolution Lmax=700 and Nr=270. The dotted line
shows $\pi/6$ while hatched bands indicate the positions of three simple
attractors; $\eta=0.35$.}
\label{sketch}
\end{minipage}
\hfil
\begin{minipage}[t]{.46\linewidth}
 \centerline{\includegraphics[width=1.1\linewidth]{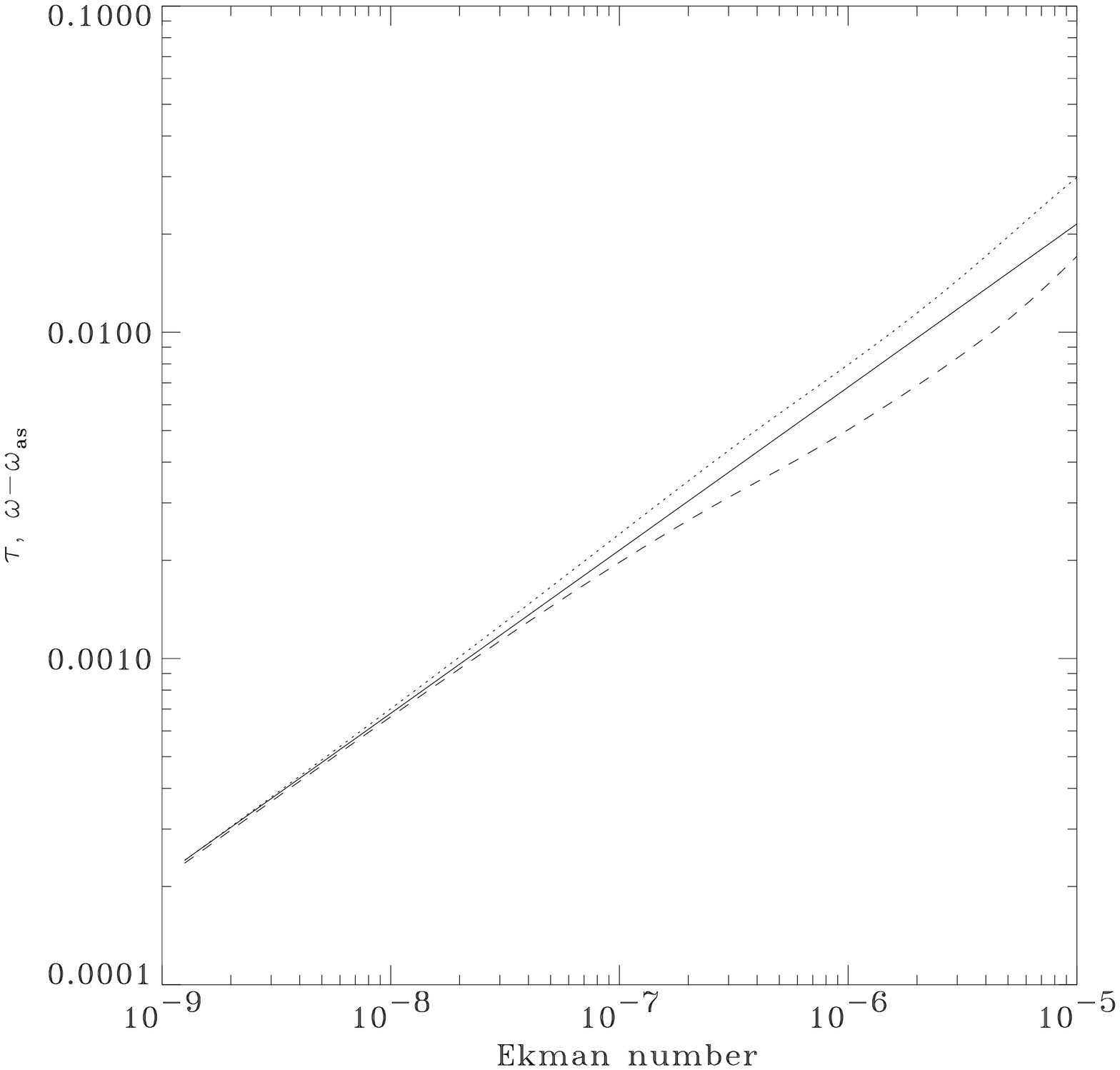}}
\caption[]{Asymptotic behaviour of the eigenvalue associated with the
eigenmode plotted in figure~\ref{num_sol}a. The dashed line represents
$\omega-\omega_{as}$ where $\omega_{as}$ is in fact $\omega_1$ given by
\eq{om_as} (cf appendix~\ref{appC}), while the dotted line is for the
damping rate. The solid line represents the `theoretical' law
$E^{1/2}$.} 
\label{as_laws}
\end{minipage}
\hfil
\end{figure}

For a given upper bound of the damping rate, eigenvalues
will be packed around the roots $\omega_i$ and around the allowed
frequencies of the set $\sin(p\pi/2q)$. In figure~\ref{sketch}, we
computed the distribution of least-damped eigenvalues when $E=10^{-8}$,
\ie for 140 evenly spaced frequencies between 0 and $1/\sqrt{2}$, we
computed the eight least-damped modes. This figure offers a glimpse at
the asymptotic distribution of eigenvalues in the complex plane: we
clearly see three main bands\footnote{They are $[0.3959,0.4162]$,
$[0.5290,0.5554]$ and $[0.6,0.6266]$; the first and third are
illustrated by attractors in figure~\ref{num_sol}; the second is illustrated in
\cite{RGV00b}; for all $\eta=0.35$.} of attractors where modes are more
damped and the two frequencies $\sin(\pi/6)$ and $\sin(\pi/4)$ where
least-damped modes tend to accumulate; the $\sin(\pi/4)$ case is conspicuous.
Note also the similarity with figure~\ref{asymp_spectrum} where the
three bands made by the aforementioned attractors are also clearly
visible.


From the asymptotic behaviour \eq{as_lyap} of the Lyapunov exponent in the
vicinity of the roots $\omega_i$, we can derive that 

\[ \omega = \omega_i + aE^{2-6\sigma} + \cdots \]

\noi while an order of magnitude evaluation of the ratio of dissipation to
kinetic energy yields the asymptotic law of the damping rate $\tau$ 

\[ \tau  = -b E^{1-2\sigma} + \cdots \]

\noi This asymptotic behaviour of eigenvalues is best illustrated in
figure~\ref{as_laws} where\linebreak \mbox{$\sigma=1/4$}. Such a mode
is the least-damped one in its frequency range and is therefore not
perturbed by other eigenvalues. This is likely the reason why it shows
its asymptotic regime at rather `high' Ekman numbers.

Concerning the eigenmodes, it is worth noting that \eq{as_lyap} and
\eq{phi_lamda} imply that the spacing of adjacent rays scales like
$E^{1/4}$ (if $\sigma=1/4$); therefore, the distance between two rays
of an attractor remains the same when it is rescaled by $E^{1/4}$ and
one may conclude that each mode in the form of a viscous attractor
keeps a self-similar structure as the Ekman number vanishes.

\section{Discussion}

Ending this paper, we think that the asymptotic behaviour of inertial
modes in a spherical shell when the Ekman number vanishes can be
anticipated even if some points remain in the shadows.

We have seen at the beginning of the paper that the trajectories of
characteristics in general converge towards an attractor; exceptions are
when the sphere is full or the inner core is small enough to let
a finite number of periodic orbits remain (which are associated with
critical latitudes commensurable with $\pi$). Leaving the full sphere
for which analytical solutions exist since \cite{Bryan1888}, the generic
behaviour of characteristics is that they converge towards an attractor
which is a periodic orbit residing in some frequency band.

The knowledge of the characteristic trajectories can be used
immediately in two-dimensional problems in order to construct a solution
of the inviscid problem. This solution contains an arbitrary function
which needs to be specified on some fundamental interval(s). This makes
the eigenvalues always infinitely degenerate. In three dimensions, the
trajectories cannot be used so efficiently but their convergence towards
an attractor can be used to show the divergence of the velocity field at
zero viscosity. In two dimensions, this divergence allowed us to prove
the non-square-integrability of velocity fields associated with
attractors, implying the absence of eigenvalues in a large fraction of
the frequency band $[0,2\Omega]$.

Beside the singularities generated by attractors, we also shed new light
on the singularity arising at the critical latitude of the inner shell.
We thus generalized the result of \cite{StRic69} that the velocity field
diverges as the inverse of the square root of the distance to the
characteristic grazing the inner shell; this singularity also makes the
velocity field not square-integrable.

Among all these singular solutions, a small set of regular modes
`survive': they are purely toroidal modes which, thanks to a velocity
field which has no radial component, do not suffer the constraints imposed
by characteristics paths. Numerical computations of the whole spectrum
give a strong evidence that these modes are the only regular ones.

When viscosity is included, all the aforementioned singularities appear
in the form of shear layers. In the asymptotic r\'egime, we therefore
expect that attractors will feature the viscous solutions. However,
this asymptotic r\'egime may be reached at extremely low values of the
Ekman number, some of which may not even be relevant astrophysically or
geophysically. We may therefore face some intermediate r\'egime where
the milder singularity at critical latitude plays an important part in
featuring the amplitude of a mode.

Our numerical investigations of the structure of shear layers which are
generated by the different singularities revealed a rather complex
structure of nested layers scaling  with $E^\sigma$, $0<\sigma<1/3$
where the value $\sigma=1/4$ seems to be favoured. In
some simple cases, where the velocity field shows no node in the
transverse direction of a shear layer, numerical results indeed show a
scaling
with $E^{1/4}$. Our boundary layer analysis demonstrated that all these
internal layers should contain an inner $\sigma=\frac{1}{3}$-layer which is
similar to the vertical Stewartson layers; however, the mapping made by
characteristics influences the scales larger than $E^{1/3}$ and
therefore makes a local analysis insufficient to determine the structure
of the outer parts of the layers.

The upper bound $\sigma<1/3$ has been derived using a heuristic model of
an inertial wave packet traveling along an attractor; from this model,
we also showed that the asymptotic spectrum of eigenvalues can be
derived, once the structure of the shear layer can be computed. An
important result of this analysis is that the limits of
eigenfrequencies as $E\tv 0$, do not form a dense set in $[0,2\Omega]$
contrary to the case of the full sphere.

The asymptotic behaviour of inertial modes and their associated
eigenvalues is therefore slowly becoming clearer: As the Ekman number
decreases, more and more eigenmodes are concentrated along the attractors
associated with their frequency; once this asymptotic regime is reached
the frequency of the mode changes slowly with viscosity so that the
Lyapunov exponent of the attractor decreases (in absolute value) and
converges toward the frequency where this exponent is zero. We are thus
in the position of describing r\'egimes with extremely low values of the
Ekman numbers that are relevant in astrophysics ($E=10^{-18}$ for a
radiative zone of a star) or geophysics ($E=10^{-15}$ for the liquid core
of the Earth) and which are way out of reach numerically.

 However, some points remain in the shadows, the most challenging one
being the structure of the shear layers. As we observed, this structure
builds up from a large scale phenomenon which is represented by the
mapping of characteristics and a small-scale one which is diffusion. To
the best of our knowledge, such a problem has never been investigated
in the past.  Since the three-dimensional case is much more involved
than the two dimensional case, because of the intrusion of Riemann
functions, we think that this latter case should be investigated first;
this will be the subject of future work.

\medskip

The foregoing results were derived from the analysis of inertial waves
of a fluid contained in a spherical shell but it is clear that they are of
wider generality. They can be easily generalized to any container of the
same topology, like ellipsoidal shells, or be used qualitatively for any
kind of container. In the case of ellipsoidal shells, even axial symmetry
can be relaxed: since constraints imposed by characteristic surfaces are
in a meridional plane, no small-scale should appear in the
$\varphi$-direction. Attractors are robust structures and only the
longest ones are sensitive to small modifications of the shape of the
boundaries; however, as they would transform into other long attractors, the
final solution would not be much affected. This kind of `structural
stability' is important when applying these results to real objects like
the core of the Earth which is obviously not a perfect spherical shell
\cite[see the discussion in][]{rieu99}.

Our results also naturally extend to all systems governed by a
spatially hyperbolic equations. Hence, one will find similar properties
for gravity modes \cite[]{ML95,RN99} or hydromagnetic modes
\cite[]{Malkus67}.

Now the next question raised by our results concern their implications
when the modes are of finite amplitude. These may concern the
development of the elliptic instability since this instability is precisely
an instability of inertial modes \cite[see][]{rieu99} or may affect
the transport properties of the fluid which are much enhanced around
attractors \cite[]{MBSL97,DRV99,DY99}.

In the same context, one may wonder whether the attractors can be
studied experimentally. At the moment, only one attractor has been
detected experimentally, using gravity waves of a stably stratified
fluid \cite[]{MBSL97} or inertial waves \cite[]{Maas00}. The main
obstacle for detecting attractors with experiments is the rather large
value of the Ekman number of experiments. Numerical calculations have
indeed shown that this number should not exceed a few 10$^{-8}$. Using
water in a spherical shell with a radius of 20~cm demands a rotational
speed of 12000 rpm which is difficult to achieve and raises experimental
problems. Also attractors should not be confused with other phenomena
which emphasize characteristics paths like the critical latitude
singularity or a forced perturbation like a discontinuity in velocity
forced by boundary conditions (e.g. the split disk case).

Finally, these new features of inertial modes may also have some
interesting consequences in astrophysics. It is now well-known that
rapidly rotating neutron stars can lose a substantial amount of angular
momentum when some inertial modes become unstable because of a coupling
with gravitational radiation \cite[see][]{anderss98,LOM98}. This instability
therefore controls the
rotation speed limit of neutron stars and this limit would be the higher,
the more damped are inertial modes. Stars with a core or density jump
will therefore be more stable than others, a fact which may be used to
give new constraints on the state of matter inside neutron stars
\cite[]{rieu00}.

\begin{acknowledgments}
We wish to thank Keith Aldridge and Leo Maas
for helpful discussions and a careful
reading of the manuscript.  We acknowledge support from the GdR
CNRS/IFREMER 1074 (M\'ecanique des Fluides G\'eophysique et
Astrophysiques).  Part of the calculations have been carried out on the
Cray C98 of the `Institut du D\'eveloppement et des Ressources en
Informatique Scientifique' (IDRIS)  and on the CalMip machine of the
`Centre Interuniversitaire de Calcul de Toulouse' (CICT) which are
 gratefully acknowledged.
\end{acknowledgments}

\appendix
\section{arcsin($\sqrt{3/7})$ }\label{appA}

In this appendix, we show in which cases sin$(p\pi/q)$ is the square
root of a rational.  This establishes that $\sqrt{3/7}$ is not
the sinus of any rational fraction of $\pi$.  The proof may exist
in the mathematical literature, but we have not been able to locate it;
we therefore propose here a simple proof of the result.

Let $p/q$ be a rational number, $p$ and $q$ being coprimes, and let us 
suppose that $\sin(p\pi/q)=\sqrt{a/b}$, where $a/b$ is a rational number. Then 
$\cos 2p\pi/q=1 - 2\sin^2 p\pi/q$ will be a rational number.  

Let us take first the case where $q$ is prime.  Then $\exp(2ip\pi/q)$ is
a $q^{th}$ root of unity, and as such is a solution of:

\beq
X^{q-1}+X^{q-2}+\cdots+1=0,
\eeq
provided $q > 1$. But one has also:
\beq
X^{q-1}+X^{q-2}+\cdots+1=\prod_{m=1}^{q-1} \lp X-e^{i2m\pi/q}\rp,
\label{roots}
\eeq
since the $\exp(i2m\pi/q)$, $m=1,\dots, q-1$ are roots of the polynomial
(called cyclotomic polynomial).  

Since $\exp(i2p\pi/q)$ is root of this polynomial, so is $\exp(-i2p\pi/q)$,
and they are different if $q > 2$.  The product (\ref{roots}) thus
contains 

\beq (X-\exp(i2p\pi/q))(X-\exp(-i2p\pi/q))=X^2- 2X\cos(2p\pi/q) +1
\eeqn{product}

\noi Therefore, since cos$(2p\pi/q)$ is rational,
\mbox{$X^2- 2X\cos(2p\pi/q) +1$} is a rational polynomial,
which divides
the cyclotomic polynomial $X^{q-1}+X^{q-2}+\dots+1$.  But Gauss proved
(see \cite{jacob85} p. 272) that this polynomial is irreducible in the 
field of rationals if $q$ is prime. Therefore the degree of this polynomial,
which is $q-1$, cannot exceed two, since otherwise it would be reducible.
So $q=1,2,3$ are the only possibilities.

If $q$ is not prime, then the polynomial $X^{q-1}+X^{q-2}+\dots+1$ is not
irreducible.  But $\exp(2ip\pi/q)$ is still a $q^{th}$ root of unity.
If $\exp(2ipm\pi/q)\neq 1$ for all $m < q$, one says that $\exp(2ip\pi/q)$
is a primitive $q^{th}$ root of unity.  
$\exp(i2p\pi/q)$ is a primitive $q^{th}$ root of unity since $p$ is prime
to $q$ (\cite{HW75}).  Then obviously $\exp(-i2p\pi/q)$
is a primitive $q^{th}$ root, different from $\exp(i2p\pi/q)$
since $q>2$.  Therefore the polynomial 
whose roots are all the primitive $q^{th}$ roots of unity contains the factor
\eq{product}.  But
this polynomial is irreducible in $\bbbq$ (cyclotomic polynomial) (see
\cite{HW75}, \cite{jacob85}). So if the degree of the polynomial 
is greater than two, there is a contradiction.  The degree
of the polynomial is equal to the Euler function $\phi (q)$ which is
the number of positive integers not greater than and prime to $q$.  
One has $\phi (q)> 2$ for $q>6$.  

Therefore the only possible values 
for $q$ are $q=1,2,3,4,6$.  By direct verification, one sees
that $\sqrt{3/7}$ is not sin$(p\pi/q)$ for one of these $q$.

\section{Lyapunov exponent of two attractors}\label{appC}

\subsection{Some example of Lyapunov exponents}

\subsubsection{Parameters of the equatorial attractor}

The equatorial attractor drawn in figure~\ref{orbs}a exists for all 
$\lambda$'s such that the
following equation has a root for $\phi_4 \in [0,\lambda[$, 
the latitude of the point of reflection on the inner sphere,

\[ 6\lambda = \arccos(\eta\cos(\lambda+\phi_4)) +
\arccos(\eta\cos(\lambda-\phi_4)) \]
as can be derived from the relations

\greq
\phi_1+\phi_2=2\lambda\\
\phi_2+\phi_3=-2\lambda\\
\cos(\phi_1+\lambda) = \eta\cos(\phi_4+\lambda)\\
\cos(\phi_3-\lambda) = \eta\cos(\phi_4-\lambda)
\egreqn{eqlos}

The frequency band $[\omega_1,\omega_2[$ where this attractor exists is
such that:

\beq \omega_1 = \frac{\sqrt{1-\eta}}{2}\qquad {\rm and}\qquad \omega_2=
\sin\lambda_2, \quad {\rm with}\quad \cos(6\lambda_2 - \arccos\eta)
=\eta\cos(2\lambda_2) \eeqn{om_as}

\noi If $\eta=0.35$, we find $\omega_1=0.403112887$ and $\omega_2=0.412474677$.

For this orbit the Lyapunov exponent is given by

\beq \Lambda(\omega) = -\frac{1}{4}\ln C(\omega) \eeqn{lyap4l}

\noi where $C(\omega)$ is the contraction coefficient (this orbit
is attracting when it is followed in the trigonometric sense). Using the
colatitudes of the reflection points, it can be
obtained with

\[ C=C_1C_2C_3C_4=
\left|
\frac{\sin(\phi_1-\lambda)}{\sin(\phi_1+\lambda)}
\frac{\sin(\phi_2+\lambda)}{\sin(\phi_2-\lambda)}
\frac{\sin(\phi_3-\lambda)}{\sin(\phi_3+\lambda)}
\frac{\sin(\phi_4+\lambda)}{\sin(\phi_4-\lambda)}
\right| \]

If we use the fact that $\phi_1+\phi_2=2\lambda$ and
$\phi_2+\phi_3=-2\lambda$, we finally get

\beq \Lambda(\omega) = -\frac{1}{4}\ln
\left|\frac{\sin(\phi_3-\lambda)}{\sin(\phi_3+5\lambda)}
\frac{\sin(\phi_4+\lambda)}{\sin(\phi_4-\lambda)}
\right| \eeqn{lyap_eq405}

Such a relation can also be obtained by differentiation of the formulae
relating the angles  $\phi_1$, $\phi_3$ and $\phi_4$ in the third and
fourth relations of~\eq{eqlos}; this yields:

\beq \frac{d\phi_1}{d\phi_3} =
\frac{\sin(\phi_3-\lambda)}{\sin(\phi_3+5\lambda)}
\frac{\sin(\phi_4+\lambda)}{\sin(\phi_4-\lambda)} \eeqn{lyap5}

\subsubsection{The associated polar attractor}

Reflection points of this attractor (left in figure~\ref{orbs}a)
are related by:

\greq
\phi_1+\phi_4=2\lambda \\
\phi_2+\phi_3=2\lambda \\
\phi_3+\phi_4=-2\lambda \\
\cos(\phi_2+\lambda)=\eta\cos(\phi_5+\lambda)\\
-\cos(\phi_1+\lambda)=\eta\cos(\phi_5-\lambda)\\
\egreq

\noi which implies solving

\beq 8\lambda = \arccos(\eta\cos(\phi_5+\lambda)) +
\arccos(-\eta\cos(\lambda-\phi_5)) \eeq

\noi for  $\lambda \leq \phi_5 \leq \pi/2$. The bounds of the interval
of existence are $[\lambda_1,\lambda_2]$  such that

\[ \cos(8\lambda_1 - \arccos(-\eta))=\eta\cos 2\lambda_1 \]
\[ \cos 4\lambda_2 + \eta\sin\lambda_2 = 0\]

\noi Here $\Lambda(\lambda_1) = -\infty$ , $\Lambda(\lambda_2) = 0$.
When $\eta=0.35$ we find $\omega_1=0.395915$ and $\omega_2=0.416185$.

In a similar way as we derived \eq{lyap_eq405}, we find that for the
polar orbit, the Lyapunov exponent reads

\beq \Lambda(\omega) =
-\frac{1}{5}\ln\left|\frac{\sin(\lambda+\phi_1)}{\sin(7\lambda-\phi_1)}
\frac{\sin(\lambda+\phi_5)}{\sin(\lambda-\phi_5)}\right|
\eeqn{lyap_pol405}

\subsection{$\Lambda$ in the neighbourhood of the point where $\Lambda=0$}

\subsubsection{General result}

In this subsection we prove that generically
near a critical latitude $\lambda_0$
which displays
a periodic orbit having $\Lambda=0$, the Lyapunov exponent
behaves like the square root of the variation of the critical latitude.
Actually, this is a general result valid for a one-dimensional mapping
that depends on one control parameter ($\lambda$ in our case)
near a tangent bifurcation point, provided the mapping is sufficiently smooth
there.

\begin{figure}
 \centerline{\includegraphics[width=8cm]{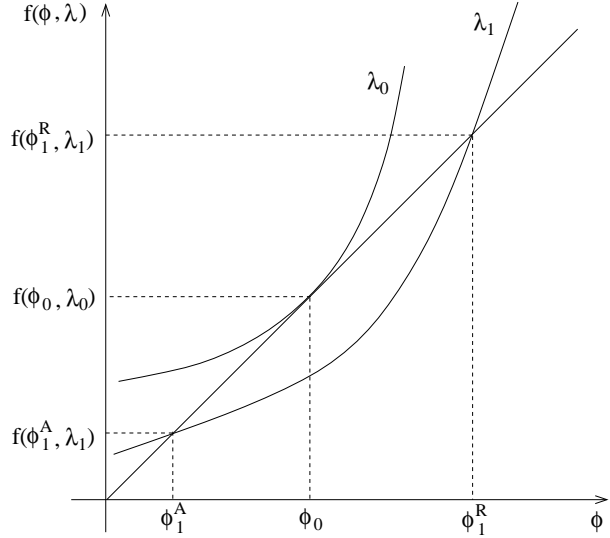}}
\caption{Generic behaviour of the mapping near a tangent bifurcation point
$(\phi_0,\lambda_0)$.
}
\label{fig:lyap0}
\end{figure}

Let $f(\phi,\lambda)$ be the mapping (for $\phi$, in our case it is
$f^N(\phi)$  with $f$ defined in \eq{applic} and $N$ is the period of
the orbit) depending on the parameter $\lambda$.
Let 
$\lambda_0$ be the value of $\lambda$ for which $\Lambda=0$, and
$\phi_0$ the fixed point of the corresponding map 
($f(\phi_0,\lambda_0)=\phi_0$).
$\lambda_0$ is a bifurcation point, as illustrated in figure \ref{fig:lyap0},
and therefore $\partial f/\partial \phi|_{\phi_0,\lambda_0}=1$.

Let $\phi_1$ be a fixed point for the mapping at $\lambda_1$ near 
$\lambda_0$:
$\phi_1=f(\phi_1,\lambda_1)$.
Let us define 
\[
f_{ij}\equiv
   \left.\frac{\partial ^{i+j}f}{\partial\phi^i\partial\lambda^j}\right|_{\phi_0,\lambda_0},
\quad \delta\phi\equiv\phi_1-\phi_0,\quad\delta\lambda\equiv\lambda_1-\lambda_0
\]
A Taylor expansion of $f$ around $(\phi_0,\lambda_0)$ up to 
order 2 yields:
\[
f(\phi_1,\lambda_1)=f(\phi_0,\lambda_0) + f_{10}\delta\phi + f_{01}\delta\lambda+
\frac{1}{2}f_{20}\delta\phi^2+
\frac{1}{2}f_{02}\delta\lambda^2+
f_{11}\delta\lambda\delta\phi+...
\]

Since $f(\phi_1,\lambda_1)=\phi_1$, 
$f(\phi_0,\lambda_0)=\phi_0$, and $f_{10}=1$, we get:
\[
f_{20}\delta\phi^2+2f_{11}\delta\lambda\delta\phi+2f_{01}\delta\lambda+
f_{02}\delta\lambda^2=0
\]
Solving this equation for $\delta\phi$ gives:
\[
\delta\phi=\frac{-f_{11}\delta\lambda\pm\sqrt{
(f_{11}^2-f_{20}f_{02})\delta\lambda^2-2f_{01}f_{20}\delta\lambda}}{f_{20}}
\]
for the coordinates of the two fixed points at $\lambda_1$, $\phi_1^A$
and $\phi_1^R$.

To leading order in $\delta\lambda$ we obtain:
\beq
\delta\phi\sim\pm\sqrt{-2\frac{f_{01}}{f_{20}}}\sqrt{\delta\lambda}
\eeqn{phi_lamda}

That is, the displacement of the fixed point $\delta\phi$ behaves like
the square root of the variation of the control parameter $\lambda$.

We compute finally the Lyapunov exponent associated with
the mapping at $(\phi_1,\lambda_1)$:
\[
\Lambda=\ln\left|\left.\frac{\partial f}{\partial\phi}\right|_{\phi_1,\lambda_1}\right|
\sim
\ln\left|f_{10}+f_{20}\delta\phi+f_{11}\delta\lambda+\dots\right|
\sim\ln\left|1\pm f_{20}\sqrt{\frac{-2f_{01}}{f_{20}}}\sqrt{\delta\lambda}\right|
\sim\pm f_{20}\sqrt{\frac{-2f_{01}}{f_{20}}}\sqrt{\delta\lambda}
\]

The positive value corresponds to the repulsive fixed point 
$\phi_1^R=\phi_0+\sqrt{-2\frac{f_{01}}{f_{20}}}\sqrt{\delta\lambda}$,
while the negative value is associated with the attractive fixed point
$\phi_1^A=\phi_0-\sqrt{-2\frac{f_{01}}{f_{20}}}\sqrt{\delta\lambda}$.
Therefore it follows that if $\Lambda(\omega_0)=0$ for an attractor,
then,

\[ \Lambda(\omega) \simeq -A\sqrt{|\omega-\omega_0|} \]
in the neighbourhood of $\omega_0$.

\subsubsection{Example of the equatorial attractor}

Let us differentiate relation \eq{lyap4l}

\[ \frac{d\Lambda}{d\omega} = -\frac{1}{4 C
\cos\lambda}\frac{dC}{d\lambda} \]

\noi where $C$ is given in \eq{lyap5} for instance. We introduce now the
new variables $\alpha=\phi_4$ and $\phi=\phi_3$ so
that $C$ reads

\[ C = \frac{\sin(\phi-\lambda)}{\sin(5\lambda+\phi)}
\frac{\sin(\alpha+\lambda)}{\sin(\alpha-\lambda)} \]

We now evaluate $\frac{dC}{d\lambda}$ at the point where $\Lambda=0$;
let us call $\lambda_0$ this point, it follows that:

\beq \frac{dC}{d\lambda}(\lambda_0) = 2\alpha' \cot\lambda_0 -
2(2+\phi')\cot3\lambda_0 \eeq

\noi where $\alpha' = d\alpha/d\lambda$ and $\phi'=d\phi/d\lambda$.
We used the fact that $\lambda_0$ verifies

\[ \cos3\lambda_0 = \eta\cos\lambda_0 \]

Using now \eq{eqlos}, we have

\greq
\cos(5\lambda+\phi) = \eta\cos(\lambda +\alpha)\\
\cos(\lambda-\phi) = \eta\cos(\lambda -\alpha)
\egreq

\noi which we differentiate with respect to $\lambda$

\greq
(5+\phi')\sin(5\lambda+\phi) = \eta(1+\alpha')\sin(\lambda
+\alpha)\\
(1-\phi')\sin(\lambda-\phi) = \eta(1-\alpha')\sin(\lambda-\alpha)
\egreq

\noi and solve that system for $\alpha'$ and $\phi'$. Its determinant
is 

\[ \Delta =
\eta\sin(\lambda-\alpha)\sin(5\lambda+\phi)(1-e^{-4\Lambda}) \]

\noi which yields

\[ \Delta \simeq 4\Lambda\eta\sin\lambda_0\sin3\lambda_0 \]

\noi in the vicinity of $\lambda_0$. Setting 

\[ \phi' = \frac{N_\phi}{\Delta},\qquad {\rm and} \qquad  \alpha' =
\frac{N_\alpha}{\Delta} \]

\noi it turns out that

\[ N_\phi(\lambda_0) = 2\eta\sin\lambda_0 f(\lambda_0) \neq 0\]
\[ N_\alpha(\lambda_0) = 2\sin3\lambda_0 f(\lambda_0) \neq 0\]

\noi with $f(\lambda) = \eta\sin\lambda-3\sin3\lambda$.
This shows that $\alpha'$ and $\phi'$ tend to infinity in $\lambda_0$.
After substitution it turns out that

\[ f(\lambda_0) = -(3+\eta)\sqrt{1-\eta} \]

\noi Hence

\[ \frac{dC}{d\lambda}(\lambda_0) = \frac{\cos\lambda_0}{\Lambda} H(\eta)
\qquad {\rm with} \qquad
H(\eta) = -\frac{4(\eta+3)}{\eta\sqrt{1-\eta}}\lc 1 -
\frac{\eta^2}{(\eta+2)^2}\rc \]

Therefore

\[  \frac{d\Lambda}{d\omega} = -\frac{H(\eta)}{4\Lambda} \]

\noi and finally
\[ \Lambda = -\sqrt{H/2}(\omega-\omega_1)^{1/2} \]

At $\omega_1$ we thus have

\[ \frac{\Lambda}{(\omega-\omega_1)^{1/2}} \tv \lc
\frac{2(\eta+3)}{\eta\sqrt{1-\eta}}\lp 1 -
\frac{\eta^2}{(\eta+2)^2}\rp \rc^{1/2}=G(\eta) \]

\noi If $\eta=0.35$, then $G(\eta)=4.8184134$.

\section{The number of attractors at a given frequency}\label{appB}

We show here that the number of periodic orbits for a given frequency
is bounded by the number of discontinuous points.

 To demonstrate this point, let us suppose the mapping has $p$
discontinuities.  This means that $f^k$ will have $kp$ discontinuities,
since the image of any discontinuous point is a discontinuous
point.  This means that $f^k$ is $C^{\infty }$ on $kp$ interval.  Each
interval is bounded by two discontinuous points.

\begin{figure}
\centerline{\includegraphics[width=7cm,angle=0]{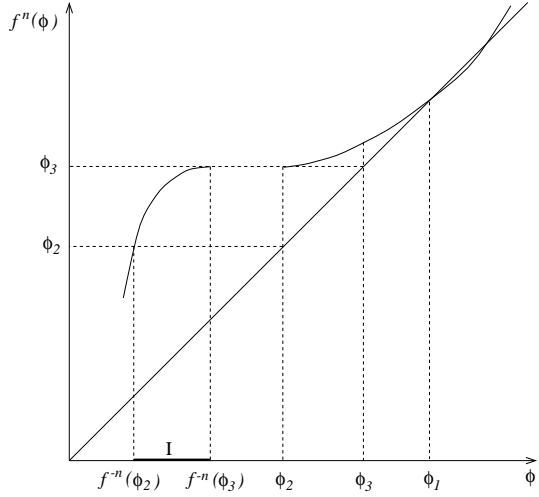}}
\caption[]{ Generic illustration of the mapping $f^n$ in the
neighbourhood of an attractive fixed point $\phi_1$. $\phi_2$ is the
point of discontinuity nearest to $\phi_1$. The interval
$I=[f^{-n}(\phi_2), f^{-n}(\phi_3)[$, after application of the mapping
$f^n$, enters the basin of attraction of $\phi_1$.} \label{intervals}
\end{figure}

When an attractor of period $n$ appears, it means that the orbit
bounces $n$ times on the outer shell. Therefore, $n$ intervals in the
graph of $f^n$ will cross the straight line $y=x$, where they will be
locked in subsequent iterations of $f^n$.

We shall denote by $\phi_1$ the attractive point closest to the end of
the interval, and by $\phi_2$ the nearest discontinuous point of $f^n$,
which bounds the locked interval; thus, $[\phi_1,\phi_2[$ belongs to
the basin of attraction of $f^n$. Now, $\phi_1$ is a fixed point of
$f^n$ and therefore of any iteration of $f^n$ or $f^{-n}$.  On the
contrary, $\phi_2$ is not.  It is a point of discontinuity of $f^n$,
therefore $f^{-n} (\phi_2)$ is a point of discontinuity of $f^{2n}$.
Thus, $f^{-n}([\phi_1,\phi_2[)$ belongs to the basin of attraction of
$f^{2n}$; since, in general, $f^{-n}(\phi_2)\neq\phi_2$,
$f^{-n}([\phi_1,\phi_2[)$ is not a continuous interval and we see that
the basin of attraction of $f^{2n}$ contains at least two intervals:
$[\phi_1,\phi_2[$ and some other interval in the neighbourhood of
$f^{-n}(\phi_2)$ (for instance $I=]f^{-n}(\phi_2),f^{-n}(\phi_3)]$ in
figure~\ref{intervals}).  Therefore at the stage $2n$, one of the new
intervals created falls into the basin.  This is true near all the $n$
attracting points of $f^n$, so $n$ additional intervals fall into the
basin at the stage $2n$.  After $n$ other iterations, $f^{-2n}(\phi_2)$
will be a new point of discontinuity of $f^{3n}$, not present in $f^n$
and $f^{2n}$, and the same argument shows that $n$ additional intervals
at least fall into the basin at stage $3n$.

This indicates that for each $n$ iteration, $n$ additional intervals
(at least) are in the basin of attraction of the attractor of period $n$.
Let us therefore consider a case with two attractors of period $n_1$ and
$n_2$; after $n_1$ iterations we get $n_1p$ intervals bounded by
discontinuities but the attractor has captured $n_1$ intervals;
therefore outside the basin of attraction of the first attractor, we
have, for the $n_1$-iterate, $n_1(p-1)$ `free' intervals at most; if the
mapping is iterated $n_1n_2$ times, we have $n_1n_2(p-1)$ free
intervals. These free intervals contain the basin of attraction of the
second attractor; but after $n_1n_2$ iterations the second attractor has
captured $n_1n_2$ intervals, therefore only $n_1n_2(p-1) -
n_1n_2=n_1n_2(p-2)$ are really free. Following this reasoning for a
third attractor, we would get $n_1n_2n_3(p-3)$ free intervals left. Thus
not more than p attractors can exist simultaneously.

It is interesting to note that numerically the number of attractors
was always found smaller or equal to $p/2$.  We note that
the argument above is actually an upper bound.  In practice, the preimage of
$[\phi_2,\phi_1]$ can include other points of discontinuity already
existing, and more than $n$ intervals can fall in the basin after each
$n$ iterations of the mapping.

The above argument is valid provided there is only one attractive point
by interval between two discontinuous points.  Actually, we never found
through extensive numerical simulations of the mapping, a case where
several attracting orbits coexist on the same interval (apart for the
values of the frequency where families of neutral orbits exist).
Actually, such a case would be related to a period-doubling bifurcation
which cannot exist in our system since one cannot cross the straight
line $y=x$ with a negative derivative.  Even if several orbits can
coexist on a given interval, this will happen at a finite iteration of
the mapping and the total number of periodic orbits will remain finite,
since the argument above shows that the number of such intervals is
finite.

\bibliography{../../biblio/bibnew}

\end{document}